\documentstyle[12pt]{article}
\begin{document}

\thispagestyle{empty}


\def\pxb{\left(\p \times \B - \B \times \p \right)}
\def\LAMBDA{\mbox{\rlap{$\raise3pt\hbox{--}$}{$\lambda$}}}
\def\rk{r_k}
\def\beq{\begin{equation}}
\def\eeq{\end{equation}}
\def\bea{\begin{eqnarray}}
\def\eea{\end{eqnarray}}
\def\nn{\nonumber}
\def\ba{\begin{array}}
\def\ea{\end{array}}
\def\0{{\mbox{\boldmath $0$}}}
\def\one{1\hskip -1mm{\rm l}}
\def\A{{\mbox{\boldmath $A$}}}
\def\B{{\mbox{\boldmath $B$}}}
\def\El{{\mbox{\boldmath $E$}}}
\def\F{{\mbox{\boldmath $F$}}}
\def\S{{\mbox{\boldmath $S$}}}

\def\a{{\mbox{\boldmath $a$}}}
\def\p{{\mbox{\boldmath $p$}}}
\def\hatp{{\widehat{\mbox{\boldmath $p$}}}}
\def\hatP{{\widehat{\mbox{\boldmath $P$}}}}
\def\vpi{{\mbox{\boldmath $\pi$}}}
\def\hatvpi{\widehat{\mbox{\boldmath $\pi$}}}
\def\r{{\mbox{\boldmath $r$}}}
\def\v{{\mbox{\boldmath $v$}}}
\def\w{{\mbox{\boldmath $w$}}}
\def\H{{\rm H}}
\def\hA{\widehat{A}}
\def\hB{\widehat{B}}
\def\ih{\frac{\i}{\hbar}}
\def\ixh{\i \hbar}
\def\ddx{\frac{\partial}{\partial x}}
\def\ddy{\frac{\partial}{\partial y}}
\def\ddz{\frac{\partial}{\partial z}}
\def\ddt{\frac{\partial}{\partial t}}
\def\vsig{{\mbox{\boldmath $\sigma$}}}
\def\Al{{\mbox{\boldmath $\alpha$}}}
\def\ho{\widehat{\cal H}_o}
\def\half{\frac{1}{2}}
\def\E{{\widehat{\cal E}}}
\def\O{{\widehat{\cal O}}}
\def\eps{\epsilon}
\def\g{\gamma}
\def\Vomeg{{\underline{\mbox{\boldmath $\Omega$}}}_s}
\def\hH{\widehat{H}}
\def\Vsig{{\mbox{\boldmath $\Sigma$}}}
\def\Nab{{\mbox{\boldmath $\nabla$}}}
\def\curl{{\rm curl}}
\def\bh{\bar{H}}
\def\th{\tilde{H}}

\def\zone{z^{(1)}}
\def\ztwo{z^{(2)}}
\def\zi{z_{\rm in}}
\def\zo{z_{\rm out}}

\def\At{{\widehat{A}(t)}}
\def\dAt{\frac{\partial {\widehat{A}(t)} }{\partial t}}
\def\sone{{\widehat{S}_1}}
\def\dsone{\frac{\partial {\widehat{S}_1} }{\partial t}}
\def\dO{\frac{\partial {\widehat{\cal O}}}{\partial t}}
\def\e{{\rm e}}
\def\ct{\widehat{\cal T}}

\section*{}
\addcontentsline{toc}{section}
{Abstract}

\begin{center}

{\LARGE\bf
Wavelength-Dependent Effects in Maxwell Optics}

\medskip
\medskip

{\em Sameen Ahmed KHAN} \\

\medskip
\medskip

khan@fis.unam.mx, ~ rohelakhan@hotmail.com \\
http://www.pd.infn.it/$\sim$khan/ \\
http://www.imsc.ernet.in/$\sim$jagan/khan-cv.html \\
Centro de Ciencias F\'{\i}sicas, \\
Universidad Nacional Aut\'onoma de M\'exico ({\bf UNAM}) \\
Apartado Postal 48-3,
Cuernavaca 62251,
Morelos,
{\bf M\'EXICO} \\

\end{center}

\medskip
\medskip

\begin{abstract}
We present a new formalism for light beam optics starting with an
exact eight-dimensional matrix representation of the Maxwell equations.
The Foldy-Wouthuysen iterative diagonalization technique is employed to
obtain a Hamiltonian description for a system with varying refractive
index.  Besides, reproducing all the traditional quasiparaxial terms,
this method leads to additional contributions, which are dependent on
the wavelength, in the optical Hamiltonian.  This alternate prescription
to obtain the aberration expansion is applied to the axially symmetric
graded index fiber.  This results in the wavelength-dependent
modifications of the paraxial behaviour and the aberration coefficients.
Furthermore it predicts a wavelength-dependent image rotation.  In the
low wavelength limit our formalism reproduces the Lie algebraic
formalism of optics.  The Foldy-Wouthuysen technique employed by us is
ideally suited for the Lie algebraic approach to optics.  The present
study further strengthens the close analogy between the various
prescription of light and charged-particle optics.  All the associated
machinery used in this formalism is described in the text and the
accompanying appendices.
\end{abstract}

\def\i{{\rm i}}

\section*{}
\addcontentsline{toc}{section}
{Contents}

\tableofcontents

\bigskip
\bigskip

\setcounter{section}{0}

\section{Introduction}
The traditional scalar wave theory of optics (including aberrations to
all orders) is based on the beam-optical Hamiltonian derived using the
Fermat's principle.  This approach is purely geometrical and works
adequately in the scalar regime.  The other approach is based on the
Helmholtz equation which is derived from the Maxwell equations.  In
this approach one takes the {\em square-root} of the Helmholtz operator
followed by an expansion of the radical~\cite{DFW,Dragt-Wave}.  This
approach works to all orders and the resulting expansion is no
different from the one obtained using the geometrical approach of the
Fermat's principle.

Another way of obtaining the aberration expansion is based on the
algebraic similarities between the Helmholtz equation and the
Klein-Gordon equation.  Exploiting this algebraic similarity the
Helmholtz equation is linearized in a procedure very similar to the one
due to Feschbach-Villars, for linearizing the  Klein-Gordon equation.
This brings the Helmholtz equation to a Dirac-like form and then
follows the procedure of the Foldy-Wouthuysen expansion used in the
Dirac electron theory.  This  approach, which uses the algebraic
machinery of quantum mechanics, was developed recently~\cite{KJS-1},
providing an alternative to the traditional {\em square-root}
procedure.  This scalar formalism gives rise to wavelength-dependent
contributions modifying the aberration coefficients~\cite{Khan-H-1}.
The algebraic machinery of this formalism is very similar to the one
used in the {\em quantum theory of charged-particle beam optics}, based
on the Dirac~\cite{JSSM}-\cite{J0} and the Klein-Gordon~\cite{KJ1}
equations respectively.  The detailed account for both of these is
available in~\cite{JK2}.  A treatment of beam optics taking into
account the anomalous magnetic moment is available
in~\cite{CJKP-1}-\cite{JK}.

As for the polarization: A systematic procedure for the passage from
scalar to vector wave optics to handle paraxial beam propagation
problems, completely taking into account the way in which the Maxwell
equations couple the spatial variation and polarization of light
waves, has been formulated by analysing the basic Poincar\'{e}
invariance of the system, and this procedure has been successfully
used to clarify several issues in Maxwell
optics~\cite{MSS-1}-\cite{SSM-2}.

In all the above approaches, the beam-optics and the polarization are
studied separately, using very different machineries.
The derivation of the Helmholtz equation from the Maxwell equations is
an approximation as one neglects the spatial and temporal derivatives
of the permittivity and permeability of the medium.  Any prescription
based on the Helmholtz equation is bound to be an approximation,
irrespective of how good it may be in certain situations.  It is very
natural to look for a prescription based fully on the Maxwell equations.
Such a prescription is sure to provide a deeper understanding of
beam-optics and polarization in a unified manner.  With this as the
chief motivation we construct a formalism starting with the Maxwell
equations in a matrix form: a single entity containing all the four
Maxwell equations.

In our approach we require an exact matrix representation of the
Maxwell equations in a medium taking into account the spatial and
temporal variations of the permittivity and permeability.  It is
{\it necessary} and {\it sufficient} to use $8 \times 8$ matrices for
such an exact representation.  This representation makes use of the
Riemann-Silberstein vector, which is described in Appendix-A.  The
derivation of the required matrix representation, and how it differs
from the numerous other ones is presented in Appendix-B.

The derived matrix representation of the Maxwell equations has a very
close algebraic correspondence with the Dirac equation.  This enables
us to apply the machinery of the Foldy-Wouthuysen expansion used in the
Dirac electron theory.  The Foldy-Wouthuysen transformation technique
is outlined in Appendix-C.  General expressions for the Hamiltonians
are derived without assuming any specific form for the refractive
index.  These Hamiltonians are shown to contain the extra
wavelength-dependent contributions which arise very naturally in our
approach.  In Section-IV we apply the general formalism to the specific
examples:
A. {\em Medium with Constant Refractive Index}.  This example
is essentially for illustrating some of the details of the machinery
used.

The other application, B. {\em Axially Symmetric Graded Index Medium}
is used to demonstrate the power of the formalism.  Two points are worth
mentioning, {\em Image Rotation}:  Our formalism gives rise to the
image rotation (proportional to the wavelength) and we have derived an
explicit relationship for the angle of the image rotation.  The other
pertains to the aberrations: In our formalism we get all the nine
aberrations permitted by the axial symmetry.  The traditional approaches
give six aberrations.  Our formalism modifies these six aberration
coefficients by wavelength-dependent contributions and also gives rise
to the remaining three permitted by the axial symmetry.  The existence
of the nine aberrations and image rotation are well-known in {\em
axially symmetric magnetic lenses}, even when treated classically.  The
quantum treatment of the same system leads to the wavelength-dependent
modifications~\cite{JK2}.  The alternate procedure for the Helmholtz
optics in~\cite{KJS-1, Khan-H-1} gives the usual six aberrations
(though modified by the wavelength-dependent contributions) and does
not give any image rotation.  These extra aberrations and the image
rotation are the exclusive outcome of the fact that the formalism
is based on a treatment starting with an exact matrix representation
of the Maxwell equations.

The traditional beam-optics is completely obtained from our approach
in the limit wavelength, $\LAMBDA \longrightarrow 0$, which we call as
the {\it traditional limit} of our formalism.  This is analogous to the
{\it classical limit} obtained by taking $\hbar \longrightarrow 0$ in
the quantum prescriptions.  The scheme of using the Foldy-Wouthuysen
machinery in this formalism is very similar to the one used in the
{\em quantum theory of charged-particle beam
optics}~\cite{JSSM}-\cite{JK}.  There  too one recovers the classical
prescriptions (Lie algebraic formalism of charged-particle beam optics,
to be precise) in the limit $\LAMBDA_0 \longrightarrow 0$, where
$\LAMBDA_0 = {\hbar}/{p_0}$ is the de Broglie wavelength and $p_0$ is
the design momentum of the system under study.

In this article we focus on the Hamiltonian description of the beam
optics, as is customary in the traditional prescriptions of beam
optics.  This also enables us to relate our formalism with the
traditional prescriptions.  The studies on the {\it evolution of the
fields} and the polarization are very much in progress.  Some of the
results in~\cite{SSM-2} have been obtained as the lowest order
approximation of the more general framework developed here.  These
shall be presented elsewhere~\cite{Khan-M-4}.


\section{Traditional Prescriptions}
Recalling, that in the traditional scalar wave theory for treating
monochromatic quasiparaxial light beam propagating along the positive
$z$-axis, the $z$-evolution of the optical wave function $\psi(\r)$ is
taken to obey the Schr\"{o}dinger-like equation
\bea
\i \LAMBDA\frac{\partial }{\partial z} \psi (\r)
= \widehat{H} \psi (\r)\,,
\label{Schr}
\eea
where the optical Hamiltonian $\widehat{H}$ is formally given by the
radical
\bea
\widehat{H} = - \left({n^2 (\r) - \hatp_\perp^2} \right)^{1/2}\,,
\eea
and $n (\r) = n (x , y , z)$ is the varying refractive index.  In beam
optics the rays are assumed to propagate almost parallel to the
optic-axis, chosen to be $z$-axis, here.  That is,
$\left| \hatp_\perp \right| \ll 1$.
The refractive index is the order of unity.
For a medium with uniform refractive index, $n (\r) = n_0$ and
the Taylor expansion of the radical is
\bea
\left({n^2 (\r) - \hatp_\perp^2} \right)^{1/2}
& = &
n_0 \left\{1 - \frac{1}{n_0^2} \hatp_\perp^2 \right\}^{1/2} \nn \\
& = &
n_0 \left\{
1 - \frac{1}{2 n_0^2} \hatp_\perp^2
- \frac{1}{8 n_0^4} \hatp_\perp^4
- \frac{1}{16 n_0^6} \hatp_\perp^6 \right. \nn \\
& & \left. \quad \qquad \qquad
- \frac{5}{128 n_0^8} \hatp_\perp^8
- \frac{7}{256 n_0^{10}} \hatp_\perp^{10} - \cdots
\right\}\,.
\eea
In the above expansion one retains terms to any desired degree of
accuracy in powers of
$\left(\frac{1}{n_0^2} \hatp_\perp^2\right)$.  In general the
refractive index is not a constant and varies.  The variation of the
refractive index $n (\r)$, is expressed as a Taylor expansion in the
spatial variables $x$, $y$ with $z$-dependent coefficients.  To get
the beam optical Hamiltonian one makes the expansion of the radical
as before, and retains terms to the desired order of accuracy in
$\left(\frac{1}{n_0^2} \hatp_\perp^2\right)$ along with all the other
terms (coming from the expansion of the refractive index $n (\r)$) in
the phase-space components up to the same order.  In this expansion
procedure the problem is partitioned into paraxial behaviour $+$
aberrations, order-by-order.

In relativistic quantum mechanics too, one has the problem of
understanding the behaviour in terms of nonrelativistic limit $+$
relativistic corrections, order-by-order.  In the Dirac theory of the
electron this is done most conveniently through the Foldy-Wouthuysen
transformation~\cite{Foldy,BD}.  The beam optical Hamiltonian derived,
starting with the exact matrix representation of the Maxwell equations
has a very close algebraic resemblance with the Dirac case, accompanied
by the analogous physical interpretations.  The details of this
correspondence and the Foldy-Wouthuysen transformation are given in
Appendix-C.

\section{The Beam-Optical Formalism}

Matrix representations of the Maxwell equations are very
well-known~\cite{Moses}-\cite{Birula}.  However, all these
representations lack an exactness or/and are given in terms of a
{\em pair} of matrix equations.  A treatment expressing the Maxwell
equations in a single matrix equation instead of a {\em pair} of matrix
equations was obtained recently~\cite{Khan-M-1}-\cite{Khan-M-3}.  This
representation contains all the four Maxwell equations in presence of
sources taking into account the spatial and temporal variations of the
permittivity $\epsilon (\r , t)$ and the permeability $\mu (\r , t)$.

Maxwell equations~\cite{Jackson,Ponofsky-Phillips} in an inhomogeneous
medium with sources are
\bea
\Nab \cdot {\mbox{\boldmath $D$}} \left(\r , t \right)
=
\rho\,, \nn \\
\Nab \times {\mbox{\boldmath $H$}} \left(\r , t \right)
- \frac{\partial }{\partial t}
{\mbox{\boldmath $D$}} \left(\r , t \right)
=
{\mbox{\boldmath $J$}}\,, \nn \\
\Nab \times {\mbox{\boldmath $E$}} \left(\r , t \right)
+
\frac{\partial }{\partial t}
{\mbox{\boldmath $B$}} \left(\r , t \right)
= 0\,, \nn \\
\Nab \cdot {\mbox{\boldmath $B$}} \left(\r , t \right)
= 0\,.
\label{Maxwell-1}
\eea
We assume the media to be linear, that is
${\mbox{\boldmath $D$}} = \epsilon (\r , t) \El$, and
${\mbox{\boldmath $B$}} = \mu (\r , t) {\mbox{\boldmath $H$}}$,
where $\epsilon$ is the {\bf permittivity of the medium} and
$\mu$ is the {\bf permeability of the medium}.
The magnitude of the velocity of light in the medium is given by
$v (\r , t)  = \left|{\mbox{\boldmath $v$}} (\r , t)\right|
= {1}/{\sqrt{\epsilon (\r , t) \mu (\r , t)}}$.  In vacuum we
have, $\epsilon_0 = 8.85 \times 10^{- 12} {C^2}/{N. m^2}$ and
$\mu_0 = 4 \pi \times 10^{- 7} {N}/{A^2}$.
Following the notation in~\cite{Birula, Khan-M-1} we use the
Riemann-Silberstein vector given by
\bea
\F^{\pm} \left(\r , t \right)
& = &
\frac{1}{\sqrt{2}}
\left(
\sqrt{\epsilon (\r , t)} \El \left(\r , t \right)
\pm \i \frac{1}{\sqrt{\mu (\r , t)}} \B \left(\r , t \right) \right)\,.
\label{R-S-Conjugate}
\eea
We further define,
\bea
\Psi^{\pm} (\r , t)
=
\left[
\ba{c}
- F_x^{\pm}  \pm \i F_y^{\pm} \\
F_z^{\pm} \\
F_z^{\pm} \\
F_x^{\pm} \pm \i F_y^{\pm}
\ea
\right]\,, \quad
W^{\pm}
=
\left(\frac{1}{\sqrt{2 \epsilon}}\right)
\left[
\ba{c}
- J_x \pm \i J_y \\
J_z - v \rho \\
J_z + v \rho \\
J_x \pm \i J_y
\ea
\right]\,,
\eea
where $W^{\pm}$ are the vectors for the sources.  Following the
notation in~\cite{Khan-M-1} the exact matrix representation of the
Maxwell equations is
\bea
& &
\frac{\partial }{\partial t}
\left[
\ba{cc}
{\mbox{\boldmath $I$}} & {\mbox{\boldmath $0$}} \\
{\mbox{\boldmath $0$}} & {\mbox{\boldmath $I$}}
\ea
\right]
\left[
\ba{cc}
\Psi^{+} \\
\Psi^{-}
\ea
\right]
-
\frac{\dot{v} (\r , t)}{2 v (\r , t)}
\left[
\ba{cc}
{\mbox{\boldmath $I$}} & {\mbox{\boldmath $0$}} \\
{\mbox{\boldmath $0$}} & {\mbox{\boldmath $I$}}
\ea
\right]
\left[
\ba{cc}
\Psi^{+} \\
\Psi^{-}
\ea
\right] \nn \\
& & \qquad \qquad \quad
+
\frac{\dot{h} (\r , t)}{2 h (\r , t)}
\left[
\ba{cc}
{\mbox{\boldmath $0$}} & \i \beta \alpha_y \\
\i \beta \alpha_y & {\mbox{\boldmath $0$}}
\ea
\right]
\left[
\ba{cc}
\Psi^{+} \\
\Psi^{-}
\ea
\right]
\nn \\
& & \qquad \quad
=  - v (\r , t)
\left[
\ba{ccc}
\left\{
{\mbox{\boldmath $M$}} \cdot \Nab
+
{\mbox{\boldmath $\Sigma$}} \cdot {\mbox{\boldmath $u$}}
\right\}
& &
- \i \beta
\left({\mbox{\boldmath $\Sigma$}} \cdot {\mbox{\boldmath $w$}}\right)
\alpha_y
\\
- \i \beta
\left({\mbox{\boldmath $\Sigma$}}^{*} \cdot {\mbox{\boldmath $w$}}\right)
\alpha_y
& &
\left\{
{\mbox{\boldmath $M$}}^{*} \cdot \Nab
+
{\mbox{\boldmath $\Sigma$}}^{*} \cdot {\mbox{\boldmath $u$}}
\right\}
\ea
\right]
\left[
\ba{cc}
\Psi^{+} \\
\Psi^{-}
\ea
\right] \nn \\
& & \qquad \quad \quad
- \left[
\ba{cc}
{\mbox{\boldmath $I$}} & {\mbox{\boldmath $0$}} \\
{\mbox{\boldmath $0$}} & {\mbox{\boldmath $I$}}
\ea
\right]
\left[
\ba{c}
W^{+} \\
W^{-}
\ea
\right]\,,
\eea
where $(\ )^{*}$ denotes {\it complex-conjugation},
$\dot{v} = \frac{\partial v}{\partial t}$ and
$\dot{h} = \frac{\partial h}{\partial t}$.
The various matrices are
\bea
M_x
& = &
\left[
\ba{cc}
{\mbox{\boldmath $0$}} & \one \\
\one & {\mbox{\boldmath $0$}}
\ea
\right]\,, \qquad
M_y
=
\left[
\ba{cc}
{\mbox{\boldmath $0$}} & - \i \one \\
\i \one & {\mbox{\boldmath $0$}}
\ea
\right]\,, \qquad
M_z
=
\beta
=
\left[
\ba{cc}
\one & {\mbox{\boldmath $0$}} \\
{\mbox{\boldmath $0$}} & - \one
\ea
\right]\,, \nn \\
{\mbox{\boldmath $\Sigma$}}
& = &
\left[
\ba{cc}
{\mbox{\boldmath $\sigma$}} & {\mbox{\boldmath $0$}} \\
{\mbox{\boldmath $0$}} & {\mbox{\boldmath $\sigma$}}
\ea
\right]\,, \qquad
{\mbox{\boldmath $\alpha$}}
=
\left[
\ba{cc}
{\mbox{\boldmath $0$}} & {\mbox{\boldmath $\sigma$}} \\
{\mbox{\boldmath $\sigma$}} & {\mbox{\boldmath $0$}}
\ea
\right]\,, \qquad
{\mbox{\boldmath $I$}}
=
\left[
\ba{cc}
\one & {\mbox{\boldmath $0$}} \\
{\mbox{\boldmath $0$}} & \one
\ea
\right]\,,
\eea
and $\one$ is the $2 \times 2$ unit matrix.  The triplet of the Pauli
matrices, ${\mbox{\boldmath $\sigma$}}$ are
\bea
{\mbox{\boldmath $\sigma$}}
& = &
\left[
\sigma_x =
\left[
\ba{cc}
0 & 1 \\
1 & 0
\ea
\right]\,, \
\sigma_y =
\left[
\ba{lr}
0 & - \i \\
\i & 0
\ea
\right]\,, \
\sigma_z =
\left[
\ba{lr}
1 & 0 \\
0 & -1
\ea
\right]
\right]\,,
\eea
and
\bea
{\mbox{\boldmath $u$}} (\r , t)
& = &
\frac{1}{2 v (\r , t)} \Nab v (\r , t)
=
\frac{1}{2} \Nab \left\{\ln v (\r , t) \right\}
=
- \frac{1}{2} \Nab \left\{\ln n (\r , t) \right\} \nn \\
{\mbox{\boldmath $w$}} (\r , t)
& = &
\frac{1}{2 h (\r , t)} \Nab h (\r , t)
=
\frac{1}{2} \Nab \left\{\ln h (\r , t) \right\}\,.
\label{u-w}
\eea
Lastly,
\bea
{\rm Velocity ~ Function:} \, v (\r , t)
& = &
\frac{1}{\sqrt{\epsilon (\r , t) \mu (\r , t)}} \nn \\
{\rm Resistance ~ Function:} \, h (\r , t)
& = &
\sqrt{\frac{\mu (\r , t)}{\epsilon (\r , t)}}\,.
\eea
As we shall soon see, it is advantageous to use the above derived
functions instead of the permittivity, $\epsilon (\r , t)$ and the
permeability, $\mu (\r , t)$.  The functions, $v (\r , t)$ and
$h (\r , t)$ have the dimensions of velocity and resistance
respectively.

Let us consider the case without any sources~($W^{\pm} = 0$).
We further assume,
\bea
\Psi^{\pm} (\r , t)
=
\psi^{\pm} \left(\r \right) e^{- \i \omega t}\,, \qquad
\omega > 0\,,
\label{Time-Gone}
\eea
with $\dot{v} (\r , t) = 0$ and $\dot{h} (\r , t) = 0$.
Then,
\bea
& &
\left[
\ba{cc}
M_z & {\mbox{\boldmath $0$}} \\
{\mbox{\boldmath $0$}} & M_z
\ea
\right]
\frac{\partial }{\partial z}
\left[
\ba{cc}
\psi^{+} \\
\psi^{-}
\ea
\right] \nn \\
& & \quad
=
\i \frac{\omega}{v (\r)}
\left[
\ba{cc}
\psi^{+} \\
\psi^{-}
\ea
\right] \nn \\
& & \quad \quad
- v (\r)
\left[
\ba{ccc}
\left\{
{\mbox{\boldmath $M$}}_\perp \cdot \Nab_\perp
+
{\mbox{\boldmath $\Sigma$}} \cdot {\mbox{\boldmath $u$}}
\right\}
& &
- \i \beta
\left({\mbox{\boldmath $\Sigma$}} \cdot {\mbox{\boldmath $w$}}\right)
\alpha_y
\\
- \i \beta
\left({\mbox{\boldmath $\Sigma$}}^{*} \cdot {\mbox{\boldmath $w$}}\right)
\alpha_y
& &
- \left\{
{\mbox{\boldmath $M$}}_\perp^{*} \cdot \Nab_\perp
+
{\mbox{\boldmath $\Sigma$}}^{*} \cdot {\mbox{\boldmath $u$}}
\right\}
\ea
\right]
\left[
\ba{cc}
\psi^{+} \\
\psi^{-}
\ea
\right]\,. \nn \\
\label{intermediate}
\eea
At this stage we introduce the process of {\em wavization}, through
the familiar Schr\"{o}dinger replacement
\bea
- \i \LAMBDA \Nab_{\perp} \longrightarrow \hatp_{\perp}\,,
\qquad \qquad
- \i \LAMBDA \frac{\partial }{\partial z}
\longrightarrow p_z\,,
\label{wavization-1}
\eea
where $\LAMBDA = {\lambda}/{2 \pi}$ is the reduced wavelength,
$c = \LAMBDA \omega$ and $n (\r) = {c}/{v (\r)}$ is the refractive
index of the medium.  Noting, that $(pq - qp) = - \i \LAMBDA$, which is
very similar to the commutation relation, $(pq - qp) = - \i \hbar$, in
quantum mechanics.  In our formalism, `$\LAMBDA$' plays the same role
which is played by the Planck constant, `$\hbar$' in quantum mechanics.
The traditional beam-optics is completely obtained from our formalism
in the limit $\LAMBDA \longrightarrow 0$.

Noting, that $M_z^{- 1} = M_z = \beta$, we multiply both sides of
equation~(\ref{intermediate}) by
\bea
\left[
\ba{cc}
M_z & {\mbox{\boldmath $0$}} \\
{\mbox{\boldmath $0$}} & M_z
\ea
\right]^{- 1}
=
\left[
\ba{cc}
\beta & {\mbox{\boldmath $0$}} \\
{\mbox{\boldmath $0$}} & \beta
\ea
\right]
\eea
and $(\i \LAMBDA )$\,, then, we obtain
\bea
\i \LAMBDA \frac{\partial }{\partial z}
\left[
\ba{cc}
\psi^{+} (\r_\perp , z) \\
\psi^{-} (\r_\perp , z)
\ea
\right]
& = &
\widehat{\H}_g
\left[
\ba{cc}
\psi^{+} (\r_\perp , z) \\
\psi^{-} (\r_\perp , z)
\ea
\right]\,.
\label{G-H}
\eea
This is the basic optical equation, where
\bea
\widehat{\H}_g
& = &
- n_0
\left[
\ba{cc}
\beta & {\mbox{\boldmath $0$}} \\
{\mbox{\boldmath $0$}} & - \beta
\ea
\right]
+ \widehat{\cal E}_g + \widehat{\cal O}_g \nn \\
\widehat{\cal E}_g
& = &
- \left(n \left(\r \right) - n_0 \right)
\left[
\ba{cc}
\beta & {\mbox{\boldmath $0$}} \\
{\mbox{\boldmath $0$}} & \beta
\ea
\right] \beta_g \nn \\
& & \quad
+
\left[
\ba{ccc}
\beta \left\{
{\mbox{\boldmath $M$}}_\perp  \cdot \p_\perp
-  \i \LAMBDA
{\mbox{\boldmath $\Sigma$}} \cdot {\mbox{\boldmath $u$}}
\right\}
& &
{\mbox{\boldmath $0$}} \nn \\
{\mbox{\boldmath $0$}}
& &
\beta \left\{
{\mbox{\boldmath $M$}}_\perp^{*}  \cdot \p_\perp
-  \i \LAMBDA
{\mbox{\boldmath $\Sigma$}}^{*} \cdot {\mbox{\boldmath $u$}}
\right\}
\ea
\right] \nn \\
\widehat{\cal O}_g
& = &
\left[
\ba{ccc}
{\mbox{\boldmath $0$}} & &
- \LAMBDA
\left({\mbox{\boldmath $\Sigma$}} \cdot {\mbox{\boldmath $w$}}\right)
\alpha_y \\
- \LAMBDA
\left({\mbox{\boldmath $\Sigma$}}^{*} \cdot {\mbox{\boldmath $w$}}\right)
\alpha_y
& &
{\mbox{\boldmath $0$}}
\ea
\right]\,,
\label{G-Partition}
\eea
where `$g$' stands for {\em grand}, signifying the eight dimensions
and
\bea
\beta_g
=
\left[
\ba{cc}
{\mbox{\boldmath $I$}} & {\mbox{\boldmath $0$}} \\
{\mbox{\boldmath $0$}} & - {\mbox{\boldmath $I$}}
\ea
\right]\,.
\eea
The above optical Hamiltonian is exact (as exact as the Maxwell
equations in a time-independent linear media).  The approximations
are made only at the time of doing specific calculations.  Apart
from the exactness, the optical Hamiltonian is in complete algebraic
correspondence with the Dirac equation with appropriate physical
interpretations.  The relevant point is:
\bea
\beta_g \widehat{\cal E}_g
=
\widehat{\cal E}_g \beta_g\,, \qquad \quad
\beta_g \widehat{\cal O}_g
=
- \widehat{\cal O}_g \beta_g\,.
\eea
We note that the upper component ($\psi^{+}$) is coupled to the lower
component ($\psi^{-}$) through the  logarithmic divergence of the
resistance function.  If this coupling function,
${\mbox{\boldmath $w$}} = 0$, or is approximated to be zero, then the
eight dimensional equations for ($\psi^{+}$) and ($\psi^{-}$) get
completely decoupled, leading to two independent four dimensional
equations.  Each of these two equations is equivalent to the other.
These are the leading equations for our studies of beam-optics and
polarization.  In the optics context any contribution from the gradient
of the resistance function can be assumed to be negligible.  With this
reasonable assumption we can decouple the equations and reduce the
problem from eight dimensions to four dimensions.  In the following
sections we shall present a formalism with the approximation
${\mbox{\boldmath $w$}} \approx 0$.  After constructing the formalism
in four dimensions we shall also address the question of dealing with
the contributions coming from the gradient of the resistance function.
This will require the application of the Foldy-Wouthuysen
transformation technique in {\em cascade} as we shall see.  This
justifies the usage of the two derived laboratory functions in place of
permittivity and permeability respectively.

We drop the `$^{+}$' throughout, then the beam-optical Hamiltonian is
\bea
\i \LAMBDA \frac{\partial }{\partial z} \psi \left(\r \right)
& = &
\widehat{\H} \psi \left(\r \right) \nn \\
\widehat{\H}
 & = &
- n_0 \beta + \widehat{\cal E} + \widehat{\cal O} \nn \\
\widehat{\cal E}
& = &
- \left(n \left(\r \right) - n_0 \right) \beta
- \i \LAMBDA \beta
{\mbox{\boldmath $\Sigma$}} \cdot {\mbox{\boldmath $u$}} \nn \\
\widehat{\cal O}
& = &
\i \left(M_y p_x - M_x p_y \right) \nn \\
& = &
\beta \left({\mbox{\boldmath $M$}}_{\perp} \cdot \hatp_{\perp} \right)\,.
\label{H-Partition}
\eea
If we were to neglect the derivatives of the permittivity and
permeability, we would have missed the term,
$(- \i \LAMBDA \beta
{\mbox{\boldmath $\Sigma$}} \cdot {\mbox{\boldmath $u$}})$.  This
is an outcome of the exact treatment.

Proceeding with our analogy with the Dirac equation: this extra term
is analogous to the anomalous magnetic/electric moment term coupled to
the magnetic/electric field respectively in the Dirac equation.
The term we dropped (while going from the eight dimensional exact to
the four dimensional almost-exact) is analogous to the anomalous
magnetic/electric moment term coupled to the electric/magnetic fields
respectively.  However it should be borne in mind that in our exact
treatment, both the terms were derived from the Maxwell equations,
where as in the Dirac theory the anomalous terms are added based on
experimental results (some even predating the Dirac equation) and
certain arguments of invariances.  In our exact treatment of the
Maxwell optics, these are the only two terms one gets, where as in the
Dirac equation the scheme of invariances permits addition of any number
of terms!  The term,
$(- \i \LAMBDA \beta
{\mbox{\boldmath $\Sigma$}} \cdot {\mbox{\boldmath $u$}})$
is related to the polarization and we shall call it as the
{\em polarization term}.

One of the other similarities worth noting, relates to the square of
the optical Hamiltonian.
\bea
\widehat{\H}^2
& = &
\left\{
n^2 \left(\r \right)
- \hatp_{\perp}^2 \right\} - \LAMBDA^2 u^2
+ \left[
{\mbox{\boldmath $M$}}_{\perp} \cdot \hatp_{\perp} \,,
n \left(\r \right) \right] \nn \\
& & \quad
+ 2 \i \LAMBDA n (\r)
{\mbox{\boldmath $\Sigma$}} \cdot {\mbox{\boldmath $u$}}
+ \i \LAMBDA \left[
{\mbox{\boldmath $M$}}_{\perp} \cdot \hatp_{\perp} \,,
{\mbox{\boldmath $\Sigma$}} \cdot {\mbox{\boldmath $u$}}
\right] \nn \\
& = &
\left\{
n \left(\r \right)
+ \i \LAMBDA
{\mbox{\boldmath $\Sigma$}} \cdot {\mbox{\boldmath $u$}}
\right\}^2 - \hatp_{\perp}^2 \nn \\
& & \quad
+
\left[
{\mbox{\boldmath $M$}}_{\perp} \cdot \hatp_{\perp} \,,
\left\{n \left(\r \right)
+ \i \LAMBDA
{\mbox{\boldmath $\Sigma$}} \cdot {\mbox{\boldmath $u$}}
\right\} \right]\,,
\label{SQUARE}
\eea
where, $[A , B] = (AB - BA)$ is the commutator.
It is to be noted that the square of the Hamiltonian in our formalism
differs from the square of the Hamiltonian in the square-root
approaches~\cite{DFW,Dragt-Wave} and the scalar approach
in~\cite{KJS-1,Khan-H-1}.  This is essentially the same type of
difference which exists in the Dirac case.  There too, the square of
the Dirac Hamiltonian gives rise to extra pieces (such
as, $- \hbar q {\mbox{\boldmath $\Sigma$}} \cdot \B$, the
Pauli term which couples the spin to the magnetic field) which is
absent in the Schr\"{o}dinger and the  Klein-Gordon descriptions.
It is this difference in the square of the Hamiltonians which
give rise to the various extra wavelength-dependent contributions
in our formalism.  These differences persist even in the approximation
when the polarization term is neglected.

The beam optical Hamiltonian derived in~(\ref{H-Partition}) has a very
close algebraic correspondence with the Dirac equation, accompanied by
the analogous physical interpretations.  This enables us to employ
the machinery of the Foldy-Wouthuysen transformation technique.  The
details are available in Appendix-C.  To the leading order, that is to
order, $\left(\frac{1}{n_0^2} \hatp_\perp^2\right)$ the beam-optical
Hamiltonian in terms of $\E$ and $\O$ is formally given by
\bea
\i \LAMBDA \ddz \left| \psi \right\rangle
& = &
\widehat{\cal H}^{(2)} \left|\psi \right\rangle\,, \nn \\
\widehat{\cal H}^{(2)} & = &
- n_0 \beta + \E - \frac{1}{2 n_0} \beta \O^2\,.
\label{FW-2-Formal}
\eea
Note that, $\O^2 = - \hatp_\perp^2$ and
$\E = - \left(n \left(\r \right) - n_0 \right) \beta
- \i \LAMBDA \beta
{\mbox{\boldmath $\Sigma$}} \cdot {\mbox{\boldmath $u$}}$.
Since, we are primarily interested in the forward propagation,
we drop the $\beta$ from the non-matrix parts of the Hamiltonian.
The matrix terms are related to the polarization.  The formal
Hamiltonian
in~(\ref{FW-2-Formal}), expressed in terms of the phase-space
variables is:
\bea
\widehat{\cal H}^{(2)}
& = &
- \left\{n \left(\r \right)
- \frac{1}{2 n_0} \hatp_{\perp}^2 \right\}
- \i \LAMBDA \beta
{\mbox{\boldmath $\Sigma$}} \cdot {\mbox{\boldmath $u$}}\,.
\label{H-Two}
\eea
Note that one retains terms up to quadratic in the Taylor expansion
of the refractive index $n (\r)$, to be consistent with the order
of $\left(\frac{1}{n_0^2} \hatp_\perp^2\right)$.  This is the paraxial
Hamiltonian which also contains an extra matrix dependent term, which
we call as the polarization term.  Rest of it is similar to the one
obtained in the traditional approaches.

To go beyond the paraxial approximation one goes a step further in
the Foldy-Wouthuysen iterative procedure.  Note that, $\O$ is the
order of $\hatp_\perp$.  To order
$\left(\frac{1}{n_0^2} \hatp_\perp^2\right)^2$,
the beam-optical Hamiltonian in terms of $\E$ and $\O$ is formally
given by
\bea
\i \LAMBDA \ddz \left|\psi \right\rangle
& = &
\widehat{\cal H}^{(4)} \left|\psi \right\rangle\,, \nn \\
\widehat{\cal H}^{(4)}
& = &
- n_0 \beta + \E - \frac{1}{2 n_0} \beta \O^2 \nn \\
& & - \frac{1}{8 n_0^2}
\left[\O ,
\left(\left[\O , \E \right] + \i \LAMBDA \ddz \O \right) \right] \nn \\
& & + \frac{1}{8 n_0^3} \beta
\left\{
 \O^4
+
\left(\left[\O , \E \right] + \i \LAMBDA \ddz \O \right)^2
\right\}\,.
\label{FW-4-Formal}
\eea
Note that $\O^4 = \hatp_\perp^4$, and $\ddz \O = 0$.
The formal Hamiltonian in~(\ref{FW-4-Formal})
when expressed in terms of the phase-space variables is
\bea
\widehat{\cal H}^{(4)}
& = &
-
\left\{n (\r)  - \frac{1}{2 n_0} \hatp_\perp^2
- \frac{1}{8 n_0^3} \hatp_\perp^4 \right\} \nn \\
& &
- \frac{1}{8 n_0^2}
\left\{
\left[\hatp_\perp^2 \,, \, \left(n (\r) - n_0 \right) \right]_{+}
 \right. \nn \\
& &
\left. \qquad \qquad \qquad
\vphantom{\hatp_{\perp}^2}
+ 2 \left( p_x \left(n (\r) - n_0 \right) p_x
+
p_y \left(n (\r) - n_0 \right) p_y \right) \right\} \nn \\
& &
- \frac{\i}{8 n_0^2}
\left\{
\left[p_x \,, \, \left[p_y \,, \left(n (\r) - n_0 \right) \right]_{+}
\right]\,
-
\left[p_y \,, \left[p_x \,, \left(n (\r) - n_0 \right) \right]_{+}
\right]
\right\} \nn \\
& &
+ \frac{1}{8 n_0^3}
\left\{
\left[p_x \,, \left(n (\r) - n_0 \right)\right]_{+}^2
+
\left[p_y \,, \left(n (\r) - n_0 \right) \right]_{+}^2
\right\} \nn \\
& &
+ \frac{\i}{8 n_0^3}
\left\{
\left[
\left[p_x \,, \, \left(n (\r) - n_0 \right) \right]_{+} \,, \,
\left[p_y \,, \, \left(n (\r) - n_0 \right) \right]_{+}
\right]
\right\} \nn \\
& & \quad
\cdots
\label{H-Four}
\eea
where $[A , B]_{+} = (AB + BA)$ and `$\cdots$' are the contributions
arising from the presence of the polarization term.  Any further
simplification would require information about the refractive
index $n (\r)$.

Note that, the paraxial Hamiltonian~(\ref{H-Two}) and the leading order
aberration Hamiltonian~(\ref{H-Four}) differs from the ones derived in
the traditional approaches.  These differences arise by the presence of
the wavelength-dependent contributions which occur in two guises.  One
set occurs totally independent of the polarization term in the basic
Hamiltonian.  This set is a multiple of the unit matrix or at most the
matrix $\beta$.  The other set involves the contributions coming from
the polarization term in the starting optical Hamiltonian.  This gives
rise to both matrix contributions and the non-matrix contributions, as
the squares of the polarization matrices is unity.   We shall discuss
the contributions of the polarization to the beam optics elsewhere.
Here, it suffices to note existence of the the wavelength-dependent
contributions in two distinguishable guises, which are not present in
the traditional prescriptions.

\subsection{When {\bf w} $\ne$ 0}
In the previous sections we assumed, ${\mbox{\boldmath $w$}} = 0$, and
this enabled us to develop a formalism using $4 \times 4$ matrices
{\em via} the Foldy-Wouthuysen machinery.  The Foldy-Wouthuysen
transformation enables us to eliminate the odd part in the $4 \times 4$
matrices, to any desired order of accuracy.  Here too we have the
identical problem, but a step higher in dimensions.  So, we need to
apply the Foldy-Wouthuysen to reduce the strength of the odd part in
eight dimensions.  This will reduce the problem from eight to four
dimensions.

We start with the grand beam optical equation in~(\ref{G-H})
and proceed with the Foldy-Wouthuysen transformations as before,
but with each quantity in double the number of dimensions.
Symbolically this means:
\bea
& &
\widehat{\H}
\longrightarrow
\widehat{\H}_g\,, \qquad
\psi
\longrightarrow
\psi_g
=
\left[
\ba{cc}
\psi^{+} \\
\psi^{-}
\ea
\right]\,, \nn \\
& &
\widehat{\cal E}
\longrightarrow
\widehat{\cal E}_g\,, \qquad
\widehat{\cal O}
\longrightarrow
\widehat{\cal O}_g\, \nn \\
& &
n_0
\longrightarrow
n_g
=
n_0
\left[
\ba{cc}
\beta & {\mbox{\boldmath $0$}} \\
{\mbox{\boldmath $0$}} & - \beta
\ea
\right]\,.
\eea
The first Foldy-Wouthuysen iteration gives
\bea
\widehat{\cal H}_g^{(2)}
& = &
- n_0
\left[
\ba{cc}
\beta & {\mbox{\boldmath $0$}} \\
{\mbox{\boldmath $0$}} & - \beta
\ea
\right]
+ \E_g - \frac{1}{2 n_0} \beta_g \O_g^2\, \nn \\
& = &
- n_0
\left[
\ba{cc}
\beta & {\mbox{\boldmath $0$}} \\
{\mbox{\boldmath $0$}} & \beta
\ea
\right] \beta_g
+ \E_g
+ \frac{1}{2 n_0} \LAMBDA^2
{\mbox{\boldmath $w$}} \cdot {\mbox{\boldmath $w$}}
\left[
\ba{cc}
\beta & {\mbox{\boldmath $0$}} \\
{\mbox{\boldmath $0$}} & - \beta
\ea
\right] \beta_g\,.
\label{G-2-Formal}
\eea
We drop the $\beta_g$, as before and then get the following
\bea
\i \LAMBDA \frac{\partial }{\partial z} \psi \left(\r \right)
& = &
\widehat{\H} \psi \left(\r \right) \nn \\
\widehat{\H}
 & = &
- n_0 \beta + \widehat{\cal E} + \widehat{\cal O} \nn \\
\hat{\cal E}
& = &
- \left(n \left(\r \right) - n_0 \right) \beta
- \i \LAMBDA \beta
{\mbox{\boldmath $\Sigma$}} \cdot {\mbox{\boldmath $u$}}
+ wide
\frac{1}{2 n_0} \LAMBDA^2 w^2 \beta \nn \\
\widehat{\cal O}
& = &
\i \left(M_y p_x - M_x p_y \right) \nn \\
& = &
\beta \left({\mbox{\boldmath $M$}}_{\perp} \cdot \hatp_{\perp} \right)\,,
\eea
where,
$w^2 = {\mbox{\boldmath $w$}} \cdot {\mbox{\boldmath $w$}}$, the square
of the logarithmic gradient of the resistance function.  This is how
the basic beam optical Hamiltonian~(\ref{H-Partition}) gets modified.
The next degree of accuracy is achieved by going a step further in the
Foldy-Wouthuysen iteration and obtaining the $\widehat{\cal H}_g^{(4)}$.
Then, this would be the higher refined starting beam optical
Hamiltonian, further modifying the basic beam optical
Hamiltonian~(\ref{H-Partition}).  This way, we can apply the
Foldy-Wouthuysen in {\em cascade} to obtain the higher order
contributions coming from the logarithmic gradient of the resistance
function, to any desired degree of accuracy.  We are very unlikely to
need any of these contributions, but it is very much possible to keep
track of them.


\section{Applications}

In the previous sections we presented the exact matrix representation
of the Maxwell equations in a medium with varying permittivity and
permeability following the recipe in~\cite{Khan-M-1}.  From this we
derived an exact optical Hamiltonian, which was shown to be in close
algebraic analogy with the Dirac equation.  This enabled us to apply
the machinery of the Foldy-Wouthuysen transformation and we obtained an
expansion for the beam-optical Hamiltonian which works to all orders.
Formal expressions
were obtained for the paraxial Hamiltonian and the leading order
aberrating Hamiltonian, without assuming any form for the refractive
index.  Even at the paraxial level the wavelength-dependent effects
manifest by the presence of a matrix term coupled to the logarithmic
gradient of the refractive index.  This matrix term is very similar
to the spin term in the Dirac equation and we call it as
the {\em polarizing term} in our formalism.  The aberrating Hamiltonian
contains numerous wavelength-dependent terms in two guises:  One of
these is the explicit wavelength-dependent terms coming from the
commutators inbuilt in the formalism with $\LAMBDA$ playing the role
played by $\hbar$ in quantum mechanics.  The other set arises from the
the polarizing term.

Now, we apply the formalism to specific examples.  One is the medium
with constant refractive index.  This is perhaps the only problem
which can be solved exactly in a closed form expression.  This is just
to illustrate how the aberration expansion in our formalism can be
summed to give the familiar exact result.

The next example is that of the axially symmetric graded index medium.
This example enables us to demonstrate the power of the formalism,
reproducing the familiar results from the traditional approaches and
further giving rise to new results, dependent on the wavelength.


\subsection{Medium with Constant Refractive Index}
Constant refractive index is the simplest possible system.  In our
formalism, this is perhaps the only case where it is possible to do
an exact diagonalization.  This is very similar to the exact
diagonalization of the free Dirac Hamiltonian.  From the experience of
the Dirac theory we know that there are hardly any situations where one
can do the exact diagonalization.  One necessarily has to resort to
some approximate diagonalization procedure.  The Foldy-Wouthuysen
transformation scheme provides the most convenient and accurate
diagonalization to any desired degree of accuracy.
So we have adopted the Foldy-Wouthuysen scheme in our formalism.

For a medium with constant refractive index,
$n \left(\r \right) = n_c$, we have,
\bea
\widehat{\H}_{c}
& = & - n_c \beta +
\i \left(M_y p_x - M_x p_y \right)\,,
\label{H-Constant}
\eea
which is exactly diagonalized by the following transform,
\bea
T^{\pm}
& = &
\exp{\left[\i \left(\pm \i \beta \right) \O \theta \right]} \nn \\
& = &
\exp{ \left[\mp \i \beta
\left(M_y p_x - M_x p_y \right)
\theta \right]} \nn \\
& = &
\cosh \left(\left|\hatp_{\perp} \right| \theta \right)
\mp \i \frac{\beta
\left(M_y p_x - M_x p_y \right)}
{\left|\hatp_{\perp}\right|}
\sinh \left(\left|\hatp_{\perp} \right| \theta \right)\,.
\label{T-D}
\eea
We choose,
\bea
\tanh \left(2 \left|\hatp_{\perp} \right| \theta \right)
=
\frac{\left|\hatp_{\perp} \right|}{n_c}\,,
\label{tanh}
\eea
then
\bea
T^{\pm}
=
\frac{\left(n_c + P_z \right)
\mp \i \beta \left(M_y p_x - M_x p_y \right)}
{\sqrt{2 P_z \left(n_c + P_z \right)}}\,,
\eea
where $P_z = + \sqrt{\left(n_c^{2} - \hatp_{\perp}^2 \right)}$.
Then we obtain,
\bea
\widehat{\H}_c^{\rm diagonal}\,
& = &
T^{+} \widehat{\H}_{c} T^{-} \nn \\
& = &
T^{+} \left\{- n_c \beta
+ \i \left(M_y p_x - M_x p_y \right) \right\} T^{-} \nn \\
& = &
- \left\{n_c^{2} - \hatp_{\perp}^2 \right\}^{\frac{1}{2}} \beta\,.
\label{Constant-Diagonal}
\eea
We next, compare the exact result thus obtained with the approximate
ones, obtained through the systematic series procedure we have developed.
\bea
\widehat{\cal H}^{(4)}_c
& = &
- n_c
\left\{1 - \frac{1}{2 n_c^2} \hatp_{\perp}^2
- \frac{1}{8 n_c^4} \hatp_\perp^4 - \cdots \right\} \beta \nn \\
& \approx &
- n_c \left\{1 - \frac{1}{n_c^2} \hatp_\perp^2\right\}^{\frac{1}{2}} \beta \nn \\
& = &
- \left\{n_c^2 - \hatp_\perp^2 \right\}^{\frac{1}{2}} \beta \nn \\
& = &
\widehat{\H}_c^{\rm diagonal}\,.
\label{Constant-approximate}
\eea

Knowing the Hamiltonian, we can compute the transfer maps.  The
transfer operator between any pair of points
$\left\{(z^{\prime \prime} , z^{\prime}) \left|
z^{\prime \prime} \right. > z^{\prime} \right\}$
on the $z$-axis, is formally given by
\bea
\left|\psi (z^{\prime \prime} , z^{\prime}) \right|
=
\widehat{\cal T} (z^{\prime \prime} , z^{\prime})
\left|\psi (z^{\prime \prime} , z^{\prime}) \right\rangle \,,
\label{}
\eea
with
\bea
& & \i \LAMBDA \frac{\partial}{\partial z}
\widehat{\cal T} (z^{\prime \prime} , z^{\prime})
=
\widehat{\cal H} \widehat{\cal T} (z^{\prime \prime} , z^{\prime})\,,
\quad
\widehat{\cal T} (z^{\prime \prime} , z^{\prime})
=
\widehat{\cal I}\,, \nn \\
& & \nn \\
& & \widehat{\cal T} (z^{\prime \prime} , z^{\prime})
=
\wp \left\{\exp
\left[- \frac{\i}{\LAMBDA}
\int_{z^\prime}^{z^{\prime \prime}} dz\,
\widehat{\cal H} (z) \right] \right\} \nn \\
& & \quad
=
\widehat{\cal I}
- \frac{\i}{\LAMBDA} \int_{z^\prime}^{z^{\prime \prime}} dz
\widehat{\cal H} (z) \nn \\
& & \qquad
+ \left(- \frac{\i}{\LAMBDA}\right)^2
\int_{z^\prime}^{z^{\prime \prime}} dz
\int_{z^\prime}^{z} d z^\prime
\widehat{\cal H} (z) \widehat{\cal H} (z^\prime) \nn \\
& & \qquad
+ \ldots\,,
\label{Transfer-1}
\eea
where $\widehat{\cal I}$ is the identity operator and $\wp$ denotes the
path-ordered exponential.  There is no closed form expression for
$\widehat{\cal T} (z^{\prime \prime} , z^{\prime})$ for an arbitrary
choice of the refractive index $n (\r)$.  In such a situation the most
convenient form of the expression for the $z$-evolution
operator $\widehat{\cal T} (z^{\prime \prime} , z^{\prime})$, or the
$z$-propagator, is
\beq
\widehat{\cal T} (z^{\prime \prime} , z^{\prime})
=
\exp{
\left[- \frac{\i}{\LAMBDA} \widehat{T}
(z^{\prime \prime} , z^{\prime}) \right]}\,,
\label{Transfer-2}
\eeq
with
\bea
\hat {T} (z^{\prime \prime} , z^{\prime})
& = &
\int_{z^\prime}^{z^{\prime \prime}} dz \widehat{\cal H} (z) \nn \\
& & \qquad
+ \frac{1}{2} \left(- \frac{\i}{\LAMBDA} \right)
\int_{z^\prime}^{z^{\prime \prime}} dz
\int_{z^\prime}^{z}  d z^\prime
\left[\widehat{\cal H} (z)\,, \widehat{\cal H} (z^\prime) \right] \nn \\
& & \qquad
+ \ldots \,,
\label{T-Magnus}
\eea
as given by the Magnus formula~\cite{Magnus}, which is described in
Appendix-D.  We shall be needing these expressions in the next example
where the refractive index is not a constant.

Using the procedure outlined above we compute the transfer operator,
\bea
& & \widehat{U}_c \left(z_{\rm out}, z_{\rm in} \right)
= \exp{ \left[- \frac{\i}{\LAMBDA} \Delta z {\cal H}_c \right]} \nn \\
& & =
\exp{ \left[+ \frac{\i}{\LAMBDA} n_c \Delta z
\left\{1 - \frac{1}{2} \frac{\widehat{p}_{\perp}^2}{n_c^2}
- \frac{1}{8} \left(\frac{\widehat{p}_{\perp}^2}{n_c^2} \right)^2
- \cdots \right\} \right]}\,, 
\label{Constant-U}
\eea
where, $\Delta z = \left(z_{\rm out}, z_{\rm in} \right)$.
Using~(\ref{Constant-U}), we compute the transfer maps
\bea
\left(
\ba{c}
\left\langle \r_{\perp} \right\rangle \\
\left\langle \p_{\perp} \right\rangle
\ea
\right)_{\rm out}
=
\left(
\ba{ccc}
1 & & \frac{1}{\sqrt{n_c^2 - \p_{\perp}^2}} \Delta z \\
0 & & 1
\ea
\right)
\left(
\ba{c}
\left\langle \r_{\perp} \right\rangle \\
\left\langle \p_{\perp} \right\rangle
\ea
\right)_{\rm in}\,.
\label{Constant-Maps}
\eea
The beam-optical Hamiltonian is intrinsically aberrating.  Even for
the simplest situation of a constant refractive index, we have
aberrations to all orders.


\subsection{Axially Symmetric Graded Index Medium}
We just saw the treatment of the medium with a constant refractive
index.  This is perhaps the only problem which can be solved exactly
in a closed form expression.  This was just to illustrate how the
aberration expansion in our formalism can be obtained.  We now consider
the next example.  The refractive index of an axially symmetric
graded-index material can be most generally described by the following
polynomial (see, pp. 117 in~\cite{DFW})
\bea
n \left(\r \right) = n_0 + \alpha_2 (z) \r_{\perp}^2
+ \alpha_4 (z) \r_{\perp}^4 + \cdots\,,
\label{n-Dragt}
\eea
where, we have assumed the axis of symmetry to coincide
with the optic-axis, namely the $z$-axis without any
loss of generality.
We note,
\bea
\widehat{\cal E}
& = &
- \left\{
\alpha_2 (z) \r_{\perp}^2
+ \alpha_4 (z) \r_{\perp}^4 + \cdots\,, \right\} \beta
- \i \LAMBDA \beta
{\mbox{\boldmath $\Sigma$}} \cdot {\mbox{\boldmath $u$}} \nn \\
\widehat{\cal O}
& = &
\i \left(M_y p_x - M_x p_y \right) \nn \\
& = &
\beta \left({\mbox{\boldmath $M$}}_{\perp} \cdot \hatp_{\perp} \right)
\eea
where
\bea
{\mbox{\boldmath $\Sigma$}} \cdot {\mbox{\boldmath $u$}}
& = &
- \frac{1}{n_0} \alpha_2 (z)
{\mbox{\boldmath $\Sigma$}}_{\perp} \cdot \r_{\perp}
- \frac{1}{2 n_0}
\left(\frac{d }{d z} \alpha_2 (z)\right) \Sigma_z \r_\perp^2
\eea

To simplify the formal expression for the beam-optical Hamiltonian
$\widehat{\cal H}^{(4)}$ given in~(\ref{FW-4-Formal})-(\ref{H-Four}),
we make use of the following:
\bea
\left({\mbox{\boldmath $M$}}_{\perp} \cdot \hatp_{\perp} \right)^2
& = &
\hatp_{\perp}^2\,, \quad \quad
\O^2
=
- \hatp_{\perp}^2\,, \quad \quad
\ddz \O = 0\,, \nn \\
\left({\mbox{\boldmath $M$}}_{\perp} \cdot \hatp_{\perp} \right)
\r_{\perp}^2
\left({\mbox{\boldmath $M$}}_{\perp} \cdot \hatp_{\perp} \right)
& = &
\frac{1}{2}
\left(\r_{\perp}^2 \hatp_{\perp}^2
+ \hatp_{\perp}^2 \r_{\perp}^2 \right)
+ 2 \LAMBDA \beta \widehat{L}_z
+ 2 \LAMBDA^2\,,
\eea
where, $\widehat{L}_z$ is the angular momentum.
Finally, the beam-optical Hamiltonian to order
$\left(\frac{1}{n_0^2} \hatp_\perp^2\right)^2$ is
\bea
\widehat{\cal H}
& = &
\widehat{H}_{0\,, p}
+ \widehat{H}_{0\,, (4)} \nn \\
& & \quad
+ \widehat{H}_{0\,, (2)}^{(\LAMBDA)}
+ \widehat{H}_{0\,, (4)}^{(\LAMBDA)} \nn \\
& & \quad
+ \widehat{H}^{(\LAMBDA, \sigma)} \nn \\
\widehat{H}_{0\,, p}
& = &
- n_0 + \frac{1}{2 n_0} \hatp_{\perp}^2
- \alpha_2 (z) \r_{\perp}^2 \nn \\
\widehat{H}_{0\,, (4)}
& = &
\frac{1}{8 n_0^3} \hatp_{\perp}^4 \nn \\
& &
- \frac{\alpha_2 (z)}{4 n_0^2} \left(\r_{\perp}^2 \hatp_{\perp}^2
+ \hatp_{\perp}^2 \r_{\perp}^2 \right) \nn \\
& &
- \alpha_4 (z) \r_{\perp}^4 \nn \\
\widehat{H}_{0\,, (2)}^{(\LAMBDA)}
& = &
- \frac{\LAMBDA^2}{2 n_0^2} \alpha_2 (z)
- \frac{\LAMBDA}{2 n_0^2} \alpha_2 (z) \widehat{L}_z
+ \frac{\LAMBDA^2}{2 n_0^3} \alpha_2^2 (z) \r_{\perp}^2 \nn \\
\widehat{H}_{0\,, (4)}^{(\LAMBDA)}
& = &
\frac{\LAMBDA}{4 n_0^3} \alpha_2^2 (z)
\left(\r_{\perp}^2 \widehat{L}_z
+ \widehat{L}_z \r_{\perp}^2 \right)
+
\frac{\LAMBDA^2}{2 n_0^3} \alpha_2 (z) \alpha_4 (z) \r_\perp^4 \nn \\
\widehat{H}^{(\LAMBDA, \sigma)}
& = &
\frac{\i \LAMBDA^3}{2 n_0^3}
\left\{\frac{d }{d z} \alpha_2 (z) \right\}
\beta \Sigma_z \nn \\
& & \qquad
+
\frac{\i \LAMBDA^2}{4 n_0^3}
\alpha_2 (z)
\left(\Sigma_x p_y - \Sigma_y p_x \right) \nn \\
& & \qquad
+
\frac{\i \LAMBDA^3}{2 n_0^3}
\left\{\frac{d }{d z} \alpha_2 (z) \right\}
\Sigma_z \widehat{L}_z  \nn \\
& & \qquad wide
+ \frac{\i \LAMBDA}{4 n_0^3}
\alpha_2 (z)
\beta
\left[
{\mbox{\boldmath $\Sigma$}}_\perp \cdot \r_\perp ,
\hatp_{\perp}^2 \right]_{+} \nn \\
& & \qquad
+ \frac{\i \LAMBDA}{8 n_0^3}
\left\{\frac{d }{d z} \alpha_2 (z) \right\}
\beta \Sigma_z
\left[\r_\perp^2 , \hatp_{\perp}^2 \right]_{+} \nn \\
& & \qquad
+ \cdots
\label{H-Fiber}
\eea
where $[A , B]_{+} = (AB + BA)$ and `$\cdots$' are the numerous other
terms arising from the polarization term.  We have retained only the
leading order of such terms above for an illustration.  All these
matrix terms, related to the polarization will be addressed elsewhere.

The reasons for partitioning the beam-optical Hamiltonian
$\widehat{\cal H}$ in the above manner are as follows.  The paraxial
Hamiltonian, $\widehat{H}_{0\,, p}$, describes the ideal behaviour.
$\widehat{H}_{0\,, (4)}$ is responsible for the third-order aberrations.
Both of these Hamiltonians are modified by the wavelength-dependent
contributions given in $\widehat{H}_{0\,, (2)}^{(\LAMBDA)}$ and
$\widehat{H}_{0\,, (4)}^{(\LAMBDA)}$ respectively.  Lastly, we have
$\widehat{H}^{(\LAMBDA, \sigma)}$, which is associated with the
polarization and shall be examined elsewhere.

\subsubsection{Image Rotation}

From these sub-Hamiltonians we make several observations:

The term $\frac{\LAMBDA}{2 n_0^2} \alpha_2 (z) \widehat{L}_z$ which
contributes to the paraxial Hamiltonian, gives rise to
an {\em image rotation} by an angle $\theta (z)$:
\bea
\theta (z^{\prime \prime} , z^{\prime})
=
\frac{\LAMBDA}{2 n_0^2}
\int_{z^\prime}^{z^{\prime \prime}} d z \alpha_2 (z)\,.
\label{theta}
\eea


This image rotation (which need not be small) has no
analogue in the {\em square-root approach}~\cite{DFW,Dragt-Wave}
and the {\em scalar approach}~\cite{KJS-1,Khan-H-1}.

\subsubsection{Aberrations}
The Hamiltonian $\widehat{H}_{0\,, (4)}$ is the one we have in the
traditional prescriptions and is responsible for the six aberrations.
$\widehat{H}_{0\,, (4)}^{(\LAMBDA)}$ modifies the above six aberrations
by wavelength-dependent contributions and further gives rise to the
remaining three aberrations permitted by the axial symmetry.
Before proceeding further we enumerate all the nine
aberrations permitted by the axial symmetry.
The axial symmetry permits {\em exactly} nine third-order
aberrations which are:

\bigskip

\begin{tabular}{lll}
Symbol & Polynomial & Name \\
$C$ & $\hatp_{\perp}^4$ & Spherical Aberration \\
$K$ & $\left[\hatp_{\perp}^2 \,, \left(\hatp_\perp \cdot \r_\perp
+ \r_\perp \cdot \hatp_\perp \right) \right]_{+}$
& Coma \\
$k$ & $\hatp_{\perp}^2 \widehat{L}_z$ & Anisotropic Coma \\
$A$ & $\left(\hatp_\perp \cdot \r_\perp
+ \r_\perp \cdot \hatp_\perp \right)^{2}$ & Astigmatism \\
$a$ & $\left(\hatp_\perp \cdot \r_\perp
+ \r_\perp \cdot \hatp_\perp \right) \widehat{L}_z$
& Anisotropic Astigmatism \\
$F$ & $\left(\hatp_{\perp}^2 \r_{\perp}^2
+ \r_{\perp}^2 \hatp_{\perp}^2 \right)$ & Curvature of Field \\
$D$ & $\left[\r_{\perp}^2 \,, \left(\hatp_\perp \cdot \r_\perp
+ \r_\perp \cdot \hatp_\perp \right) \right]_{+}$
& Distortion \\
$d$ & $\r_{\perp}^2 \widehat{L}_z$ & Anisotropic Distortion \\
$E$ & $\r_{\perp}^4$ & Nameless? or POCUS
\end{tabular}

~\\
The name {\it POCUS} is used in~\cite{DFW} on page~137.

The axial symmetry allows only the terms (in the Hamiltonian)
which are produced out of, $\hatp_{\perp}^2$, $\r_{\perp}^2$,
$\left(\hatp_\perp \cdot \r_\perp + \r_\perp \cdot \hatp_\perp \right)$
and $\widehat{L}_z$.  Combinatorially, to fourth-order one would get
ten terms including $\widehat{L}_z^2$.  We have listed nine of them in
the table above.  The tenth one namely,
\bea
\widehat{L}_z^2
=
\frac{1}{2}
\left(\hatp_{\perp}^2 \r_{\perp}^2
+ \r_{\perp}^2 \hatp_{\perp}^2 \right)
-
\frac{1}{4}
\left(\hatp_\perp \cdot \r_\perp
+ \r_\perp \cdot \hatp_\perp \right)^{2}
+
\LAMBDA^2
\eea
So, $\widehat{L}_z^2$ is not listed separately.  Hence, we have only
nine third-order aberrations permitted by axial symmetry, as stated
earlier.

The paraxial transfer maps are given by
\bea
\left(
\ba{c}
\left\langle \r_{\perp} \right\rangle \\
\left\langle \p_{\perp} \right\rangle
\ea
\right)_{\rm out}
=
\left(
\ba{cc}
P & Q \\
R & S
\ea
\right)
\left(
\ba{c}
\left\langle \r_{\perp} \right\rangle \\
\left\langle \p_{\perp} \right\rangle
\ea
\right)_{\rm in}\,,
\label{Paraxial-Maps}
\eea
where~$P$, $Q$, $R$ and $S$ are the solutions of the paraxial
Hamiltonian~(\ref{H-Fiber}). The symplecticity condition tells us that
$P S - Q R = 1$.  In this particular case from the structure of the
paraxial equations, we can further conclude that: $R = P^\prime$ and
$S = Q^\prime$ where $(\ )^\prime$ denotes the $z$-derivative.

The transfer operator is most accurately and neatly expressed in terms
of the paraxial solutions, $P$, $Q$, $R$ and $S$, {\em via} the
{\em interaction picture} of the Lie algebraic formulation of light
beam optics and charged-particle beam optics~\cite{Interaction}.
\bea
\widehat{\cal T} \left(z\,, z_0\right)
& = &
\exp {
\left[
- \frac{\i}{\LAMBDA} \widehat{T}
\left(z\,, z_0\right) \right] }\,, \nn \\
& = &
\exp
\left[
- \frac{\i}{\LAMBDA}
\left\{
C \left(z^{\prime \prime}\,, z^\prime \right) \hatp_{\perp}^4
\phantom{\frac{\i}{\LAMBDA}} \right. \right. \nn \\
& & \quad \quad
+
K \left(z^{\prime \prime}\,, z^\prime \right)
\left[\hatp_{\perp}^2 \,, \left(\hatp_\perp \cdot \r_\perp
+ \r_\perp \cdot \hatp_\perp \right) \right]_{+} \nn \\
& & \quad \quad
+
k \left(z^{\prime \prime}\,, z^\prime \right)
\hatp_{\perp}^2 \widehat{L}_z \nn \\
& & \quad \quad
+ A \left(z^{\prime \prime}\,, z^\prime \right)
\left(\hatp_\perp \cdot \r_\perp
+ \r_\perp \cdot \hatp_\perp \right)^{2} \nn \\
& & \quad \quad
+ a \left(z^{\prime \prime}\,, z^\prime \right)
\left(\hatp_\perp \cdot \r_\perp
+ \r_\perp \cdot \hatp_\perp \right) \widehat{L}_z \nn \\
& & \quad \quad
+
F \left(z^{\prime \prime}\,, z^\prime \right)
\left(\hatp_{\perp}^2 \r_{\perp}^2
+ \r_{\perp}^2 \hatp_{\perp}^2 \right) \nn \\
& & \quad \quad
+
D \left(z^{\prime \prime}\,, z^\prime \right)
\left[\r_{\perp}^2 \,, \left(\hatp_\perp \cdot \r_\perp
+ \r_\perp \cdot \hatp_\perp \right) \right]_{+} \nn \\
& & \quad \quad
+
d \left(z^{\prime \prime}\,, z^\prime \right)
\r_{\perp}^2 \widehat{L}_z \nn \\
& & \quad \quad \left. \left.
+
E \left(z^{\prime \prime}\,, z^\prime \right)
\r_{\perp}^4
\vphantom{\frac{\i}{\LAMBDA}}
\right\} \right]\,.
\eea
The nine aberration coefficients are given by,
\bea
C \left(z^{\prime \prime}\,, z^\prime \right)
& = &
\int_{z^\prime}^{z^{\prime \prime}} d z
\left\{
\frac{1}{8 n_0^3} S^4
- \frac{\alpha_2 (z)}{2 n_0^2} Q^2 S^2
- \alpha_4 (z) Q^4 \right. \nn \\
& & \left. \qquad \qquad
+ \frac{\LAMBDA^2}{2 n_0^3} \alpha_2 (z) \alpha_4 (z) Q^4
\right\} \nn \\
K \left(z^{\prime \prime}\,, z^\prime \right)
& = &
\int_{z^\prime}^{z^{\prime \prime}} d z
\left\{
\frac{1}{8 n_0^3} R S^3
- \frac{\alpha_2 (z)}{4 n_0^2} QS(PS + QR)
- \alpha_4 (z) PQ^3 \right. \nn \\
& & \left. \qquad \qquad
+ \frac{\LAMBDA^2}{2 n_0^3} \alpha_2 (z) \alpha_4 (z) PQ^3
\right\} \nn \\
k \left(z^{\prime \prime}\,, z^\prime \right)
& = &
\frac{\LAMBDA}{2 n_0^3}
\int_{z^\prime}^{z^{\prime \prime}} d z
\alpha_2^2 (z) Q^2 \nn \\
A \left(z^{\prime \prime}\,, z^\prime \right)
& = &
\int_{z^\prime}^{z^{\prime \prime}} d z
\left\{
\frac{1}{8 n_0^3} R^2 S^2
- \frac{\alpha_2 (z)}{2 n_0^2} PQRS
- \alpha_4 (z) P^2 Q^2 \right.  \nn \\
& & \left. \qquad  \qquad
+ \frac{\LAMBDA^2}{2 n_0^3} \alpha_2 (z) \alpha_4 (z) P^2 Q^2
\right\} \nn \\
a \left(z^{\prime \prime}\,, z^\prime \right)
& = &
\frac{\LAMBDA}{2 n_0^3}
\int_{z^\prime}^{z^{\prime \prime}} d z
\alpha_2^2 (z) P Q \nn \\
F \left(z^{\prime \prime}\,, z^\prime \right)
& = &
\int_{z^\prime}^{z^{\prime \prime}} d z
\left\{
\frac{1}{8 n_0^3} R^2 S^2
- \frac{\alpha_2 (z)}{4 n_0^2} (P^2 S^2 + Q^2 R^2)
- \alpha_4 (z) P^2 Q^2 \right. \nn \\
& & \left. \qquad \qquad
+ \frac{\LAMBDA^2}{2 n_0^3} \alpha_2 (z) \alpha_4 (z) P^2 Q^2
\right\} \nn \\
D \left(z^{\prime \prime}\,, z^\prime \right)
& = &
\int_{z^\prime}^{z^{\prime \prime}} d z
\left\{
\frac{1}{8 n_0^3} R^3 S
- \frac{\alpha_2 (z)}{4 n_0^2} PR (PS + QR)
- \alpha_4 (z) P^3 Q \right. \nn \\
& & \left. \qquad \qquad
+ \frac{\LAMBDA^2}{2 n_0^3} \alpha_2 (z) \alpha_4 (z) P^3 Q
\right\} \nn \\
d \left(z^{\prime \prime}\,, z^\prime \right)
& = &
\frac{\LAMBDA}{2 n_0^3}
\int_{z^\prime}^{z^{\prime \prime}} d z
\alpha_2^2 (z) P^2 \nn \\
E \left(z^{\prime \prime}\,, z^\prime \right)
& = &
\int_{z^\prime}^{z^{\prime \prime}} d z
\left\{
\frac{1}{8 n_0^3} R^4
- \frac{\alpha_2 (z)}{2 n_0^2} P^2 R^2
- \alpha_4 (z) P^4 \right. \nn \\
& & \left. \qquad \qquad
+ \frac{\LAMBDA^2}{2 n_0^3} \alpha_2 (z) \alpha_4 (z) P^4
\right\}\,.
\label{fiber-aberration-coefficients}
\eea

Thus we see that the current approach gives rise to all the nine
permissible aberrations.  The six aberrations, familiar from the
traditional prescriptions get modified by the wavelength-dependent
contributions.  The extra three ($k$, $a$ and $d$, all anisotropic!)
are all pure wavelength-dependent aberrations and totally absent in the
traditional {\em square-root approach}~\cite{DFW,Dragt-Wave} and the
recently developed {\em scalar approach}~\cite{KJS-1,Khan-H-1}.
A detailed account on the classification of aberrations is available
in~\cite{KBW-1}-\cite{KBW-4}.

\section{Polarization}

Let there be Light! (with or/and without polarization)~\cite{Khan-M-4}.

\section{Concluding Remarks}
We have developed an exact matrix representation of the Maxwell
equations taking into account the spatial and temporal variations of
the permittivity and permeability.  This representation, using
$8 \times 8$ matrices is the basis for an exact formalism of Maxwell
optics
presented here.  The exact beam optical Hamiltonian, derived from this
representation has an algebraic structure in direct correspondence with
the Dirac equation of the electron.  We exploit this correspondence to
adopt the standard machinery, namely the Foldy-Wouthuysen
transformation technique of the Dirac theory, to the beam optical
formalism.  This enabled us to obtain a systematic procedure to obtain
the aberration expansion from the beam-optical Hamiltonian to any
desired degree of accuracy.  We further get the wavelength-dependent
contributions at each order, starting with the lowest-order paraxial
paraxial Hamiltonian.  Formal expressions were obtained for the
paraxial and leading order aberrating Hamiltonians, without making any
assumption on the form of the varying refractive index.

The beam-optical Hamiltonians also have the wavelength-dependent
matrix terms which are associated with the polarization.  In this
approach we have been able to derive a Hamiltonian which contains both
the beam-optics and the polarization.

In Section-IV, we applied the formalism to the specific examples and
saw how the beam-optics (paraxial behaviour and the aberrations) gets
modified by the wavelength-dependent contributions.  First of the two
examples is the {\em medium with a constant refractive index}.  This
is perhaps the only problem which can be solved exactly, in a closed
form expression.  This example is primarily for illustrating certain
aspects of the machinery we have used.

The second, and the much more interesting example is that of the
{\em axially symmetric graded index medium}.  For this example, in
the traditional approaches one gets only six aberrations.  In our
formalism we get all the nine aberrations permitted by the axial
symmetry.  The six aberration coefficients of the traditional
approaches get modified by the wavelength-dependent contributions.
It is very interesting to note that apart from the wavelength-dependent
modifications of the aberrations, this approach also gives rise to the
{\it image rotation}.  This image rotation is proportional to the
wavelength and we have derived an explicit relationship for the angle
in~(\ref{theta}).  Such, an image rotation has no analogue/counterpart
in any of the traditional prescriptions.  It would be worthwhile to
experimentally look for the predicted image rotation.  The existence of
the nine aberrations and image rotation are well-known in {\em axially
symmetric magnetic electron lenses}, even when treated classically.
The quantum treatment of the same system leads to the
wavelength-dependent modifications~\cite{JK2}.

The optical Hamiltonian has two components: {\em Beam-Optics} and
{\em Polarization} respectively.  We have addressed the former in some
detail and the later is in progress.  The formalism initiated in this
article provides a natural framework for the study of light
polarization.  This would provide a unified treatment for the
beam-optics and the polarization.  It also promises a possible
generalization of the {\em substitution} result in~\cite{SSM-2}.  We
shall present this approach elsewhere~\cite{Khan-M-4}.

The close analogy between geometrical optics and charged-particle beam
optics has been known for too long a time.  Until recently it was
possible to see this analogy only between the geometrical optics and
the classical prescriptions of charge-particle optics.  A quantum
theory of charged-particle optics was presented in recent
years~\cite{JSSM}-\cite{CJKP-1}.  With the  current development of the
non-traditional prescriptions of Helmholtz optics~\cite{KJS-1,Khan-H-1}
and the matrix formulation of Maxwell optics presented here, using the
rich algebraic machinery of quantum mechanics it is now possible to see
a parallel of the analogy at each level.  The non-traditional
prescription of the Helmholtz optics is in close
analogy with the quantum theory of charged-particles based on the
Klein-Gordon equation.  The matrix formulation of Maxwell optics
presented here is in close analogy with the quantum theory of
charged-particles based on the Dirac equation~\cite{Analogy-ICFA}.
The parallel of these analogies is described in Appendix-E.

An important omission in the present study is the study of the
{\it evolution of the fields}, $(\El\,, \B)$, which we shall address in
detail elsewhere~\cite{Khan-M-4}.  Even without the discussion of the
fields (as is the case in  several other prescriptions) the present
study is complete at the Hamiltonian level.  We have presented an
alternate and exact way of deriving the beam optical Hamiltonian, which
reproduces the established results.  Furthermore we have derived the
extra wavelength-dependent contributions.  In the low wavelength limit
our formalism reproduces the Lie algebraic formalism of optics.  The
Foldy-Wouthuysen technique employed by us is ideally suited for the Lie
algebraic approach to optics.  The present study further strengthens
the close analogy between the various prescription of light and
charged-particle beam optics~\cite{Analogy-ICFA}.

\newpage
\setcounter{section}{0}

\section*{}
\addcontentsline{toc}{section}
{Appendix A. \\
Riemann-Silberstein Vector}


\renewcommand{\theequation}{A.{\arabic{equation}}}
\setcounter{equation}{0}

\begin{center}

{\Large\bf
Appendix-A \\
Riemann-Silberstein Vector
} \\

\end{center}

The Riemann-Silberstein complex vector~\cite{Birula},
$\F \left(\r , t \right)$ built from the
electric field ${\mbox{\boldmath $D$}} \left(\r , t \right)$ and the
magnetic filed ${\mbox{\boldmath $B$}} \left(\r , t \right)$ is
given by
\bea
\F \left(\r , t \right)
& = &
\frac{1}{\sqrt{2}}
\left(
\frac{1}{\sqrt{\epsilon (\r)}} {\mbox{\boldmath $D$}} \left(\r , t \right)
+ \i \frac{1}{\sqrt{\mu (\r)}} \B \left(\r , t \right) \right) \nn \\
& = &
\frac{1}{\sqrt{2}}
\left(
\sqrt{\epsilon (\r)} \El \left(\r , t \right)
+ \i \frac{1}{\sqrt{\mu (\r)}} \B \left(\r , t \right) \right)\,,
\label{R-S-V}
\eea
where $\epsilon (\r)$ is the {\bf permittivity of the medium} and
$\mu (\r)$ is the {\bf permeability of the medium}.  In vacuum we
have $\epsilon_0 = 8.85 \times 10^{- 12} {C^2}/{N. m^2}$ and
$\mu_0 = 4 \pi \times 10^{- 7} {N}/{A^2}$.
The Riemann-Silberstein complex vector, $\F \left(\r , t \right)$
can also be derived from the potential
${\mbox{\boldmath $Z$}} \left(\r , t \right)$,
(for example, see~\cite{Ponofsky-Phillips}),
\bea
\F \left(\r , t \right)
& = &
\Nab \times
\left\{
\frac{\i}{v} \frac{\partial }{\partial t}
{\mbox{\boldmath $Z$}} \left(\r , t \right)
+
\Nab \times {\mbox{\boldmath $Z$}} \left(\r , t \right)
\right\}\,.
\eea
${\mbox{\boldmath $Z$}} \left(\r , t \right)$ is the {\em superpotential}
and is commonly known as the {\em polarization potential} or the
{\em Hertz Vector} (see~\cite{Ponofsky-Phillips}).
This further leads to the  wave-equation
\bea
\left\{
\Nab^2
- \frac{1}{v^2} \frac{\partial^2 }{\partial t^2}
\right\}
{\mbox{\boldmath $Z$}} \left(\r , t \right)
& = &
0\,.
\eea

Riemann-Silberstein vector can be used to express many of
the quantities associated with the electromagnetic field:
\bea
{\rm Poynting ~ Vector:} \, {\mbox{\boldmath $S$}}
& = &
\frac{1}{\mu} \El \times \B \nn \\
& = &
- \i v \left(\F^{\dagger} \times \F \right) \nn \\
{\rm Energy ~ Density:} \, u
& = &
\frac{1}{2}
\left(
\epsilon \El \cdot \El + \frac{1}{\mu} \B \cdot \B
\right) \nn \\
& = &
\F^{\dagger} \cdot \F \nn \\
{\rm Momentum ~ Density:} \,
{\mbox{\boldmath $p$}}_{ EB}
& = &
\epsilon \left(\El \times \B \right) \nn \\
& = &
- \frac{\i}{v} \left(\F^{\dagger} \times \F \right) \nn \\
{\rm Angular ~ Momentum ~ Density:} \, {\mbox{\boldmath {\cal $L$}}}_{ EB}
& = &
\epsilon \left\{\r \times \left(\El \times \B \right) \right\} \nn \\
& = &
- \frac{\i}{v} \left\{\r \times \left(\F^{\dagger} \times \F \right) \right\}\,.
\label{RS-quantities-1}
\eea
And
\bea
{\rm Total ~ Energy:} \, E
& = &
\frac{1}{2}
\int d^3 r \, \left\{\epsilon \El \cdot \El
+ \frac{1}{\mu} \B \cdot \B \right\} \nn \\
& = &
\int d^3 r \, \left\{\F^{\dagger} \cdot \F \right\} \nn \\
{\rm Total ~ Momentum:} \, {\mbox{\boldmath $P$}}
& = &
\epsilon \int d^3 r \, \left\{\El \times \B \right\} \nn \\
& = &
- \frac{\i}{v}
\int d^3 r \, \left\{\F^{\dagger} \times \F \right\} \nn \\
{\rm Total ~ Angular ~ Momentum:} \, {\mbox{\boldmath $M$}}
& = &
\epsilon
\int d^3 r \, \left\{\r \times \left(\El \times \B \right) \right\} \nn \\
& = &
- \frac{\i}{v}
\int d^3 r \, \left\{\r \times \left(\F^{\dagger} \times \F \right) \right\} \nn \\
{\rm Moment ~ of ~ Energy:} \, {\mbox{\boldmath $N$}}
& = &
\frac{1}{2}
\int d^3 r \, \left\{\r \left(\epsilon \El \cdot \El
+ \frac{1}{\mu} \B \cdot \B \right) \right\} \nn \\
& = &
\int d^3 r \, \left\{\r \left(\F^{\dagger} \cdot \F \right) \right\}
\label{RS-quantities-2}
\eea
In this form these quantities look like the quantum-mechanical
expectation values!  The use of the Riemann-Silberstein vector as a
possible candidate for the {\em photon wavefunction} has been advocated
for a long time~\cite{Birula}.

\newpage

\section*{}
\addcontentsline{toc}{section}
{Appendix B. \\
An Exact Matrix Representation of the Maxwell Equations}

%
\renewcommand{\theequation}{B.{\arabic{equation}}}
\renewcommand{\thesection}{B.{\arabic{section}}}
\renewcommand{\thesubsection}{B.{\arabic{subsection}}}
\setcounter{subsection}{0}
\setcounter{equation}{0}

\begin{center}

{\Large\bf
Appendix-B \\
An Exact Matrix Representation of the Maxwell Equations in a Medium
} \\

\end{center}

Matrix representations of the Maxwell equations are very
well-known~\cite{Moses}-\cite{Birula}.  However, all these
representations lack an exactness or/and are given in terms of a
{\em pair} of matrix equations.  Some of these representations are
in free space.  Such a representation is an approximation in a
medium with space- and time-dependent permittivity
$\epsilon (\r , t)$ and permeability $\mu (\r , t)$ respectively.
Even this approximation is often expressed through a pair of equations
using $3 \times 3$ matrices: one for the curl and one for the
divergence which occur in the Maxwell equations.  This practice of
writing the divergence condition separately is completely avoidable by
using $4 \times 4$ matrices for Maxwell equations in
free-space~\cite{Moses}.  A single equation using $4 \times 4$ matrices
is necessary and sufficient when $\epsilon (\r , t)$ and $\mu (\r , t)$
are treated as `local' constants~\cite{Moses,Birula}.

A treatment taking into account the variations of
$\epsilon (\r , t)$ and $\mu (\r , t)$ has been presented
in~\cite{Birula}.  This treatment uses the
Riemann-Silberstein vectors, $\F^{\pm} \left(\r , t \right)$
to reexpress the Maxwell equations as four equations: two
equations are for the curl and two are for the divergences and
there is mixing in
$\F^{+} \left(\r , t \right)$ and $\F^{-} \left(\r , t \right)$.
This mixing is very neatly expressed through the two derived
functions of $\epsilon (\r , t)$ and $\mu (\r , t)$.
These four equations are then expressed as a pair of matrix
equations using $6 \times 6$ matrices: again one for the curl and
one for the divergence.   Even though this treatment is exact it
involves a pair of matrix equations.

Here, we present a treatment which enables us to express the
Maxwell equations in a single matrix equation instead of a {\em pair}
of matrix equations.  Our approach is a logical continuation of the
treatment in~\cite{Birula}.  We use the linear combination of the
components of the  Riemann-Silberstein vectors,
$\F^{\pm} \left(\r , t \right)$ and the final matrix representation
is a single equation using $8 \times 8$ matrices.  This representation
contains all the four Maxwell equations in presence of sources taking
into account the spatial and temporal variations of the permittivity
$\epsilon (\r , t)$ and the permeability $\mu (\r , t)$.

In Section-I we shall summarize the treatment for a homogeneous medium
and introduce the required functions and notation.  In Section-II we
shall present the matrix representation in an inhomogeneous medium,
in presence of sources.

\subsection{Homogeneous Medium}
We shall start with the Maxwell
equations~\cite{Jackson, Ponofsky-Phillips} in an inhomogeneous medium
with sources,
\bea
\Nab \cdot {\mbox{\boldmath $D$}} \left(\r , t \right)
=
\rho\,, \nn \\
\Nab \times {\mbox{\boldmath $H$}} \left(\r , t \right)
- \frac{\partial }{\partial t}
{\mbox{\boldmath $D$}} \left(\r , t \right)
=
{\mbox{\boldmath $J$}}\,, \nn \\
\Nab \times {\mbox{\boldmath $E$}} \left(\r , t \right)
+
\frac{\partial }{\partial t}
{\mbox{\boldmath $B$}} \left(\r , t \right)
= 0\,, \nn \\
\Nab \cdot {\mbox{\boldmath $B$}} \left(\r , t \right)
= 0\,.
\label{ME-Maxwell-1}
\eea
We assume the media to be linear, that is
${\mbox{\boldmath $D$}} = \epsilon \El$, and
${\mbox{\boldmath $B$}} = \mu {\mbox{\boldmath $H$}}$,
where $\epsilon$ is the {\bf permittivity of the medium} and
$\mu$ is the {\bf permeability of the medium}.  In general
$\epsilon = \epsilon (\r , t)$ and $\mu = \mu (\r , t)$.  In
this section we treat them as `local' constants in the various
derivations.  The magnitude of the velocity of light in the medium
is given by
$v (\r , t)  = \left|{\mbox{\boldmath $v$}} (\r , t)\right|
= {1}/{\sqrt{\epsilon (\r , t) \mu (\r , t)}}$.  In vacuum we
have, $\epsilon_0 = 8.85 \times 10^{- 12} {C^2}/{N. m^2}$ and
$\mu_0 = 4 \pi \times 10^{- 7} {N}/{A^2}$.

One possible way to obtain the required matrix representation is
to use the Riemann-Silberstein vector~\cite{Birula} given by
\bea
\F^{+} \left(\r , t \right)
& = &
\frac{1}{\sqrt{2}}
\left(
\sqrt{\epsilon (\r , t)} \El \left(\r , t \right)
+ \i \frac{1}{\sqrt{\mu (\r , t)}} \B \left(\r , t \right) \right) \nn \\
\F^{-} \left(\r , t \right)
& = &
\frac{1}{\sqrt{2}}
\left(
\sqrt{\epsilon (\r , t)} \El \left(\r , t \right)
- \i \frac{1}{\sqrt{\mu (\r , t)}} \B \left(\r , t \right) \right)\,.
\label{ME-R-S-Conjugate}
\eea
For any homogeneous medium it is equivalent to use either
$\F^{+} \left(\r , t \right)$ or $\F^{-} \left(\r , t \right)$.  The
two differ by the sign before `$\i$' and are not the complex conjugate
of one another.  We have not assumed any form for $\El (\r , t)$ and
$\B (\r , t)$.  We will be needing both of them in an inhomogeneous
medium, to be considered in detail in Section-II.

If for a certain medium $\epsilon (\r , t)$ and $\mu (\r , t)$ are
constants (or can be treated as `local' constants under certain
approximations), then the vectors $\F^{\pm} \left(\r , t \right)$
satisfy
\bea
\i \frac{\partial }{\partial t} \F^{\pm} \left(\r , t \right)
& = &
\pm v \Nab \times \F^{\pm} \left(\r , t \right)
- \frac{1}{\sqrt{2 \epsilon}} (\i {\mbox{\boldmath $J$}}) \nn \\
\Nab \cdot \F^{\pm} \left(\r , t \right)
& = &
\frac{1}{\sqrt{2 \epsilon}} (\rho)\,.
\label{ME-RS-F-Equations-Approximate}
\eea
Thus, by using the Riemann-Silberstein vector it has been possible
to reexpress the four Maxwell equations (for a medium with constant
$\epsilon$ and $\mu$) as two equations.  The first one contains the
the two Maxwell equations with curl and the second one contains the
two Maxwell with divergences.  The first of the two equations
in~(\ref{ME-RS-F-Equations-Approximate}) can be immediately converted
into a $3 \times 3$ matrix representation.  However, this
representation does not contain the divergence conditions (the first
and the fourth Maxwell equations) contained in the second equation
in~(\ref{ME-RS-F-Equations-Approximate}).  A further compactification is
possible only by expressing the Maxwell equations in a $4 \times 4$
matrix representation.  To this end,  using the components of the
Riemann-Silberstein vector, we define,
\bea
\Psi^{+} (\r , t)
& = &
\left[
\ba{c}
- F_x^{+}  + \i F_y^{+} \\
F_z^{+} \\
F_z^{+} \\
F_x^{+} + \i F_y^{+}
\ea
\right]\,, \quad
\Psi^{-} (\r , t)
=
\left[
\ba{c}
- F_x^{-}  - \i F_y^{-} \\
F_z^{-} \\
F_z^{-} \\
F_x^{-} - \i F_y^{-}
\ea
\right]\,.
\eea
The vectors for the sources are
\bea
W^{+}
& = &
\left(\frac{1}{\sqrt{2 \epsilon}}\right)
\left[
\ba{c}
- J_x + \i J_y \\
J_z - v \rho \\
J_z + v \rho \\
J_x + \i J_y
\ea
\right]\,, \quad
W^{-}
=
\left(\frac{1}{\sqrt{2 \epsilon}}\right)
\left[
\ba{c}
- J_x - \i J_y \\
J_z - v \rho \\
J_z + v \rho \\
J_x - \i J_y
\ea
\right]\,.
\label{ME-Potentials-s}
\eea
Then we obtain
\bea
\frac{\partial}{\partial t}
\Psi^{+}
& = &
- v
\left\{ {\mbox{\boldmath $M$}} \cdot \Nab \right\} \Psi^{+}
- W^{+}\, \nn \\
\frac{\partial}{\partial t}
\Psi^{-}
& = &
- v
\left\{ {\mbox{\boldmath $M$}}^{*} \cdot \Nab \right\} \Psi^{-}
- W^{-}\,,
\label{ME-Spinorial}
\eea
where $(\ )^{*}$ denotes complex-conjugation and the triplet,
$
{\mbox{\boldmath $M$}} = \left(M_x\,, M_y\,, M_z \right)
$
is expressed in terms of
\bea
\Omega
& = &
\left[
\ba{cc}
{\mbox{\boldmath $0$}} & - \one \\
\one & {\mbox{\boldmath $0$}}
\ea
\right]\,, \qquad
\beta =
\left[
\ba{cccc}
\one & {\mbox{\boldmath $0$}} \\
{\mbox{\boldmath $0$}} & - \one
\ea
\right]\,, \qquad
\one =
\left[
\ba{cc}
1 & 0 \\
0 & 1
\ea
\right]\,.
\eea
Alternately, we may use the matrix $J = - \Omega$.  Both differ by
a sign.  For our purpose it is fine to use either $\Omega$ or $J$.
However, they have a different meaning: $J$ is {\em contravariant}
and $\Omega$ is {\em covariant}; The matrix $\Omega$ corresponds
to the Lagrange brackets of classical mechanics and $J$ corresponds
to the Poisson brackets.  An important relation is $\Omega = J^{-1}$.
The $M$-matrices are:
\bea
M_x & = &
\left[
\ba{cccc}
0 & 0 & 1 & 0 \\
0 & 0 & 0 & 1 \\
1 & 0 & 0 & 0 \\
0 & 1 & 0 & 0
\ea
\right] = - \beta \Omega  \,, \nn \\
M_y & = &
\left[
\ba{cccc}
0 & 0 & - \i & 0 \\
0 & 0 & 0 & - \i \\
\i & 0 & 0 & 0 \\
0 & \i & 0 & 0
\ea
\right] = \i \Omega \,, \nn \\
M_z & = &
\left[
\ba{cccc}
1 & 0 & 0 & 0 \\
0 & 1 & 0 & 0 \\
0 & 0 & -1 & 0 \\
0 & 0 & 0 & - 1
\ea
\right] = \beta \,.
\label{ME-M-Matrices}
\eea
Each of the four Maxwell equations are easily obtained from the
matrix representation in~(\ref{ME-Spinorial}).  This is done by taking
the sums and differences of row-I with row-IV and row-II with row-III
respectively.  The first three give the $y$, $x$ and $z$ components
of the curl and the last one gives the divergence conditions present
in the evolution equation~(\ref{ME-RS-F-Equations-Approximate}).

It is to be noted that the matrices~${\mbox{\boldmath $M$}}$ are all
non-singular and all are hermitian.  Moreover, they satisfy the usual
algebra of the Dirac matrices, including,
\bea
M_x \beta = - \beta M_x\,, \nn \\
M_y \beta = - \beta M_y\,, \nn \\
M_x^2 = M_y^2 = M_z^2 = I\,, \nn \\
M_x M_y = - M_y M_x = \i M_z\,, \nn \\
M_y M_z = - M_z M_y = \i M_x\,, \nn \\
M_z M_x = - M_x M_z = \i M_y\,.
\label{ME-Identities}
\eea

Before proceeding further we note the following: The pair
$(\Psi^{\pm} , {\mbox{\boldmath $M$}})$ are {\bf not} unique.
Different choices of $\Psi^{\pm}$ would give rise to different
${\mbox{\boldmath $M$}}$, such that the triplet
${\mbox{\boldmath $M$}}$ continues to to satisfy the algebra of the
Dirac matrices in~(\ref{ME-Identities}).
We have preferred $\Psi^{\pm}$ {\em via} the the Riemann-Silberstein
vector~(\ref{ME-R-S-Conjugate}) in~\cite{Birula}.  This vector has
certain
advantages over the other possible choices.   The Riemann-Silberstein
vector is well-known in classical electrodynamics and has certain
interesting properties and uses~\cite{Birula}.

In deriving the above $4 \times 4$ matrix representation of the
Maxwell equations we have ignored the spatial and temporal derivatives
of $\epsilon (\r , t)$ and $\mu (\r , t)$ in the first two of the
Maxwell equations.  We have treated
$\epsilon$ and $\mu$ as `local' constants.

\subsection{Inhomogeneous Medium}
In the previous section we wrote the evolution equations for the
Riemann-Silberstein vector in~(\ref{ME-RS-F-Equations-Approximate}),
for
a medium, treating $\epsilon (\r , t)$ and $\mu (\r , t)$ as `local'
constants.  From these pairs of equations we wrote the matrix form of
the Maxwell equations.  In this section we shall write the exact
equations taking into account the spatial and temporal variations
of $\epsilon (\r , t)$ and $\mu (\r , t)$.  It is very much possible to
write the required evolution equations using $\epsilon (\r , t)$ and
$\mu (\r , t)$.  But we shall follow the procedure in~\cite{Birula} of
using the two derived {\em laboratory functions}
\bea
{\rm Velocity ~ Function:} \, v (\r , t)
& = &
\frac{1}{\sqrt{\epsilon (\r , t) \mu (\r , t)}} \nn \\
{\rm Resistance ~ Function:} \, h (\r , t)
& = &
\sqrt{\frac{\mu (\r , t)}{\epsilon (\r , t)}}\,.
\eea
The function, $v (\r , t)$ has the dimensions of velocity and the
function, $h (\r , t)$ has the dimensions of resistance (measured
in Ohms).  We can equivalently use the {\em Conductance Function},
$\kappa (\r , t) = {1}/{h (\r , t)} =
{\epsilon (\r , t)}/{\mu (\r , t)}$ (measured in Ohms$^{- 1}$
or Mhos!) in place of the resistance function, $h (\r , t)$.  These
derived functions enable us to understand the dependence of the
variations more transparently~\cite{Birula}.  Moreover the derived
functions are the ones which are measured experimentally.  In terms
of these functions,  $\epsilon = {1}/{\sqrt{v h}}$ and
$\mu = \sqrt{{h}/{v}}$.
Using these functions the exact equations satisfied by
$\F^{\pm} \left(\r , t \right)$ are
\bea
\i \frac{\partial }{\partial t} \F^{+} \left(\r , t \right)
& = &
v (\r , t) \left( \Nab \times \F^{+} \left(\r , t \right) \right)
+ \frac{1}{2}
\left(\Nab v (\r , t) \times \F^{+} \left(\r , t \right) \right) \nn \\
& &
+ \frac{v (\r , t)}{2 h (\r)}
\left(\Nab h (\r , t) \times \F^{-} \left(\r , t \right) \right)
- \frac{\i}{\sqrt{2}} \sqrt{v (\r , t) h (\r , t)}\,
{\mbox{\boldmath $J$}} \nn \\
& &
+ \frac{\i}{2} \frac{\dot{v} (\r , t)}{v (\r , t)}
\F^{+} \left(\r , t \right)
+ \frac{\i}{2} \frac{\dot{h} (\r , t)}{h (\r , t)}
\F^{-} \left(\r , t \right) \nn \\
\i \frac{\partial }{\partial t} \F^{-} \left(\r , t \right)
& = &
- v (\r , t) \left( \Nab \times \F^{-} \left(\r , t \right) \right)
- \frac{1}{2}
\left(\Nab v (\r , t) \times \F^{-} \left(\r , t \right) \right) \nn \\
& &
- \frac{v (\r , t)}{2 h (\r , t)}
\left(\Nab h (\r , t) \times \F^{+} \left(\r , t \right) \right)
- \frac{\i}{\sqrt{2}} \sqrt{v (\r , t) h (\r , t)}\,
{\mbox{\boldmath $J$}} \nn \\
& &
+ \frac{\i}{2} \frac{\dot{v} (\r , t)}{v (\r , t)}
\F^{-} \left(\r , t \right)
+ \frac{\i}{2} \frac{\dot{h} (\r , t)}{h (\r , t)}
\F^{+} \left(\r , t \right) \nn \\
\Nab \cdot \F^{+} \left(\r , t \right)
& = &
\frac{1}{2 v (\r , t)}
\left(\Nab v (\r , t) \cdot \F^{+} \left(\r , t \right) \right) \nn \\
& &
+
\frac{1}{2 h (\r , t)}
\left(\Nab h (\r , t) \cdot \F^{-} \left(\r , t \right) \right) \nn \\
& &
+ \frac{1}{\sqrt{2}} \sqrt{v (\r , t) h (\r , t)}\, \rho \,, \nn \\
\Nab \cdot \F^{-} \left(\r , t \right)
& = &
\frac{1}{2 v (\r , t)}
\left(\Nab v (\r , t) \cdot \F^{-} \left(\r , t \right) \right) \nn \\
& &
+
\frac{1}{2 h (\r , t)}
\left(\Nab h (\r , t) \cdot \F^{+} \left(\r , t \right) \right) \nn \\
& &
+ \frac{1}{\sqrt{2}} \sqrt{v (\r , t) h (\r , t)}\, \rho \,,
\label{ME-RS-F-Equations-Exact}
\eea
where
$\dot{v} = \frac{\partial v}{\partial t}$ and
$\dot{h} = \frac{\partial h}{\partial t}$.
The evolution equations in~(\ref{ME-RS-F-Equations-Exact}) are exact
(for a linear media) and the dependence on the variations
of $\epsilon (\r , t)$ and $\mu (\r , t)$ has been neatly expressed
through the two derived functions.   The  {\em coupling} between
$\F^{+} \left(\r , t \right)$ and $\F^{-} \left(\r , t \right)$.
is {\em via} the gradient and time-derivative of only one derived
function namely, $h (\r , t)$ or equivalently $\kappa (\r , t)$.
Either of these can be used and both are the directly measured
quantities.  We further note that the dependence of the coupling
is logarithmic
\bea
\frac{1}{h (\r , t)} \Nab h (\r , t)
=
\Nab \left\{\ln {\left(h (\r , t) \right)} \right\}\,, \quad
\frac{1}{h (\r , t)} \dot{h} (\r , t)
=
\frac{\partial }{\partial t}
\left\{\ln {\left(h (\r , t) \right)} \right\}\,,
\label{ME-Logarithm}
\eea
where `$\ln$' is the natural logarithm.

The coupling can be best summarized by expressing the equations
in~(\ref{ME-RS-F-Equations-Exact}) in a (block) matrix form.
For this we introduce the following logarithmic function
\bea
{\cal L} (\r , t)
& = &
\frac{1}{2}
\left\{
\one \ln {\left(v (\r , t)\right)}
+
\sigma_x
\ln {\left(h (\r , t)\right)}
\right\}\,,
\label{ME-Logarithmic-Function}
\eea
where
$\sigma_x$ is one the triplet of the Pauli matrices
\bea
{\mbox{\boldmath $\sigma$}}
& = &
\left[
\sigma_x =
\left[
\ba{cc}
0 & 1 \\
1 & 0
\ea
\right]\,, \
\sigma_y =
\left[
\ba{lr}
0 & - \i \\
\i & 0
\ea
\right]\,, \
\sigma_z =
\left[
\ba{lr}
1 & 0 \\
0 & -1
\ea
\right]
\right]\,.
\eea
Using the above notation the matrix form of the
equations in~(\ref{ME-RS-F-Equations-Exact}) is
\bea
\i
\left\{
\one \frac{\partial }{\partial t}
- \frac{\partial }{\partial t} {\cal L}
\right\}
\left[
\ba{cc}
\F^{+} \left(\r , t \right) \\
\F^{-} \left(\r , t \right)
\ea
\right]
& = &
v (\r) \sigma_z
\left\{
\one \Nab + \Nab {\cal L}
\right\} \times
\left[
\ba{cc}
\F^{+} \left(\r , t \right) \\
\F^{-} \left(\r , t \right)
\ea
\right]\, \nn \\
& &
- \frac{\i}{\sqrt{2}} \sqrt{v (\r , t) h (\r , t)}\,
{\mbox{\boldmath $J$}} \nn \\
\left\{
\one \Nab - \Nab {\cal L}
\right\} \cdot
\left[
\ba{cc}
\F^{+} \left(\r , t \right) \\
\F^{-} \left(\r , t \right)
\ea
\right]
& = & + \frac{1}{\sqrt{2}} \sqrt{v (\r , t) h (\r , t)}\, \rho \,,
\label{ME-Six-Evolution}
\eea
where the dot-product and the cross-product are to be
understood as
\bea
\left[
\ba{cc}
{\mbox{\boldmath $A$}} & {\mbox{\boldmath $B$}} \\
{\mbox{\boldmath $C$}} & {\mbox{\boldmath $D$}}
\ea
\right]
\cdot
\left[
\ba{c}
{\mbox{\boldmath $u$}} \\
{\mbox{\boldmath $v$}}
\ea
\right]
& = &
\left[
\ba{c}
{\mbox{\boldmath $A$}} \cdot {\mbox{\boldmath $u$}}
+
{\mbox{\boldmath $B$}} \cdot {\mbox{\boldmath $v$}} \\
{\mbox{\boldmath $C$}} \cdot {\mbox{\boldmath $u$}}
+
{\mbox{\boldmath $D$}} \cdot {\mbox{\boldmath $v$}}
\ea
\right] \nn \\
\left[
\ba{cc}
{\mbox{\boldmath $A$}} & {\mbox{\boldmath $B$}} \\
{\mbox{\boldmath $C$}} & {\mbox{\boldmath $D$}}
\ea
\right]
\times
\left[
\ba{c}
{\mbox{\boldmath $u$}} \\
{\mbox{\boldmath $v$}}
\ea
\right]
& = &
\left[
\ba{c}
{\mbox{\boldmath $A$}} \times {\mbox{\boldmath $u$}}
+
{\mbox{\boldmath $B$}} \times {\mbox{\boldmath $v$}} \\
{\mbox{\boldmath $C$}} \times {\mbox{\boldmath $u$}}
+
{\mbox{\boldmath $D$}} \times {\mbox{\boldmath $v$}}
\ea
\right]\,.
\eea
It is to be noted that the $6 \times 6$ matrices in the evolution
equations in~(\ref{ME-Six-Evolution}) are either hermitian or
antihermitian.  Any dependence on the variations of
$\epsilon (\r , t)$ and $\mu (\r , t)$ is at best `weak'.  We further
note, $\Nab \left(\ln {\left(v (\r , t) \right)} \right)
= - \Nab \left(\ln {\left(n (\r , t) \right)} \right)$ and
$\frac{\partial }{\partial t}
\left(\ln {\left(v (\r , t) \right)} \right)
= - \frac{\partial }{\partial t}
\left(\ln {\left(n (\r , t) \right)} \right)$.
In some media, the coupling may vanish
($\Nab h (\r , t) = 0$ and $\dot{h} (\r , t) = 0$) and in the same
medium the refractive index, $n (\r , t) = {c}/{v (\r , t)}$ may
vary ($\Nab n (\r , t) \ne 0$ or/and $\dot{n} (\r , t) \ne 0$).
It may be further possible to use the approximations
$\Nab \left(\ln {\left(h (\r , t) \right)} \right)
\approx 0$ and
$\frac{\partial }{\partial t}
\left(\ln {\left(h (\r , t) \right)} \right)
\approx 0$.

We shall be using the following matrices to express the exact
representation
\bea
{\mbox{\boldmath $\Sigma$}}
=
\left[
\ba{cc}
{\mbox{\boldmath $\sigma$}} & {\mbox{\boldmath $0$}} \\
{\mbox{\boldmath $0$}} & {\mbox{\boldmath $\sigma$}}
\ea
\right]\,, \qquad
{\mbox{\boldmath $\alpha$}}
=
\left[
\ba{cc}
{\mbox{\boldmath $0$}} & {\mbox{\boldmath $\sigma$}} \\
{\mbox{\boldmath $\sigma$}} & {\mbox{\boldmath $0$}}
\ea
\right]\,, \qquad
{\mbox{\boldmath $I$}}
=
\left[
\ba{cc}
\one & {\mbox{\boldmath $0$}} \\
{\mbox{\boldmath $0$}} & \one
\ea
\right]\,,
\eea
where ${\mbox{\boldmath $\Sigma$}}$ are the Dirac spin matrices and
${\mbox{\boldmath $\alpha$}}$ are the matrices used in the Dirac
equation.  Then,
\bea
& &
\frac{\partial }{\partial t}
\left[
\ba{cc}
{\mbox{\boldmath $I$}} & {\mbox{\boldmath $0$}} \\
{\mbox{\boldmath $0$}} & {\mbox{\boldmath $I$}}
\ea
\right]
\left[
\ba{cc}
\Psi^{+} \\
\Psi^{-}
\ea
\right]
-
\frac{\dot{v} (\r , t)}{2 v (\r , t)}
\left[
\ba{cc}
{\mbox{\boldmath $I$}} & {\mbox{\boldmath $0$}} \\
{\mbox{\boldmath $0$}} & {\mbox{\boldmath $I$}}
\ea
\right]
\left[
\ba{cc}
\Psi^{+} \\
\Psi^{-}
\ea
\right] \nn \\
& & \qquad \qquad \quad
+
\frac{\dot{h} (\r , t)}{2 h (\r , t)}
\left[
\ba{cc}
{\mbox{\boldmath $0$}} & \i \beta \alpha_y \\
\i \beta \alpha_y & {\mbox{\boldmath $0$}}
\ea
\right]
\left[
\ba{cc}
\Psi^{+} \\
\Psi^{-}
\ea
\right]
\nn \\
& & \qquad \quad
=  - v (\r , t)
\left[
\ba{ccc}
\left\{
{\mbox{\boldmath $M$}} \cdot \Nab
+
{\mbox{\boldmath $\Sigma$}} \cdot {\mbox{\boldmath $u$}}
\right\}
& &
- \i \beta
\left({\mbox{\boldmath $\Sigma$}} \cdot {\mbox{\boldmath $w$}}\right)
\alpha_y
\\
- \i \beta
\left({\mbox{\boldmath $\Sigma$}}^{*} \cdot {\mbox{\boldmath $w$}}\right)
\alpha_y
& &
\left\{
{\mbox{\boldmath $M$}}^{*} \cdot \Nab
+
{\mbox{\boldmath $\Sigma$}}^{*} \cdot {\mbox{\boldmath $u$}}
\right\}
\ea
\right]
\left[
\ba{cc}
\Psi^{+} \\
\Psi^{-}
\ea
\right] \nn \\
& & \qquad \quad \quad
- \left[
\ba{cc}
{\mbox{\boldmath $I$}} & {\mbox{\boldmath $0$}} \\
{\mbox{\boldmath $0$}} & {\mbox{\boldmath $I$}}
\ea
\right]
\left[
\ba{c}
W^{+} \\
W^{-}
\ea
\right]\,, 
\eea
where
\bea
{\mbox{\boldmath $u$}} (\r , t)
& = &
\frac{1}{2 v (\r , t)} \Nab v (\r , t)
=
\frac{1}{2} \Nab \left\{\ln v (\r , t) \right\}
=
- \frac{1}{2} \Nab \left\{\ln n (\r , t) \right\} \nn \\
{\mbox{\boldmath $w$}} (\r , t)
& = &
\frac{1}{2 h (\r , t)} \Nab h (\r , t)
=
\frac{1}{2} \Nab \left\{\ln h (\r , t) \right\}
\eea
The above representation contains thirteen $8 \times 8$ matrices!
Ten of these are hermitian.  The exceptional ones are the ones that
contain the three components of ${\mbox{\boldmath $w$}} (\r , t)$,
the logarithmic gradient of the resistance function.  These three
matrices, for the resistance function are antihermitian.

We have been able to express the Maxwell equations in a matrix form in
a medium with varying permittivity $\epsilon (\r , t)$ and permeability
$\mu (\r , t)$, in presence of sources.  We have been able to do so
using a single equation instead of a {\em pair} of matrix equations.
We have used $8 \times 8$ matrices and have been able to separate the
dependence of the coupling between the upper components ($\Psi^{+}$)
and the lower components ($\Psi^{-}$) through the two laboratory
functions.  Moreover the exact matrix representation has an algebraic
structure very similar to the Dirac equation.  We feel that this
representation would be more suitable for some of the studies related
to the {\em photon wave function}~\cite{Birula}.

\newpage

\section*{}
\addcontentsline{toc}{section}
{Appendix C. \\
The Foldy-Wouthusysen Representation of the Dirac Equation}

\renewcommand{\theequation}{C.{\arabic{equation}}}
\setcounter{equation}{0}

\begin{center}

{\Large\bf
Appendix-C. \\
Foldy-Wouthuysen Transformation
} \\

\end{center}

In the traditional scheme the purpose of expanding the {\em light
optics} Hamiltonian
$\widehat{H} = - \left(n^2 (\r) - \hatp_\perp^2\right)^{1/2}$ in a
series using $\left(\frac{1}{n_0^2} \hatp_\perp^2\right)$ as the
expansion parameter is to understand the propagation of the
quasiparaxial beam in terms of a series of approximations
(paraxial + nonparaxial).  Similar is the situation in the case of the
{\em charged-particle optics}.  Let us recall that in relativistic
quantum mechanics too one has a similar problem of understanding
the relativistic wave equations as the nonrelativistic approximation
plus the relativistic correction terms in the quasirelativistic regime.
For the Dirac equation (which is first order in time) this is done most
conveniently using the Foldy-Wouthuysen transformation leading to an
iterative diagonalization technique.

The main framework of the formalism of optics, used here (and in the
charged-particle optics) is based on the transformation technique of
the Foldy-Wouthuysen theory which casts the Dirac equation in a form
displaying the different interaction terms between the Dirac particle
and and an applied electromagnetic field in a nonrelativistic and
easily interpretable form (see,~\cite{Foldy},
\cite{Pryce}-\cite{Acharya}, for a
general discussion of the role of the Foldy-Wouthuysen-type
transformations in particle interpretation of relativistic wave
equations).
The suggestion to employ the Foldy-Wouthuysen Transformation technique
in the case of the Helmholtz equation was mentioned in the literature
as a remark~\cite{Fishman}.  It was only in the recent works, that this
idea was exploited to analyze the quasiparaxial approximations for
for specific beam optical system~\cite{KJS-1, Khan-H-1}.  The
Foldy-Wouthuysen technique is ideally suited for the Lie algebraic
approach to optics.  With all these plus points, the powerful and
ambiguity-free expansion, the Foldy-Wouthuysen Transformation is still
little used in optics~\cite{Khan-FW}.
In the Foldy-Wouthuysen theory the Dirac equation is decoupled through
a canonical transformation into two two-component equations: one
reduces to the Pauli equation in the nonrelativistic limit and the
other describes the negative-energy states.

Let us describe here briefly the standard Foldy-Wouthuysen theory so
that the way it has been adopted for the purposes of the above studies
in optics will be clear.
Let us consider a charged-particle of rest-mass $m_0$, charge $q$
in the presence of an electromagnetic field characterized by
$\El = - \Nab \phi
- \frac{\partial }{\partial t} {\mbox{\boldmath $A$}}$
and $\B = \Nab \times {\mbox{\boldmath $A$}}$.
Then the Dirac equation is
\bea
\i \hbar \frac{\partial}{\partial t} \Psi(\r , t)
& = &
\widehat{H}_D \Psi(\r , t)
\label{A-FW-1} \\
\widehat{H}_D
& = &
{m_0 c^2} \beta + q \phi + c \Al \cdot \widehat{\vpi} \nn \\
& = &
{m_0 c^2} \beta + \widehat{\cal E} + \widehat{\cal O} \nn \\
\widehat{\cal E}
& = &
q \phi \nn \\
\widehat{\cal O}
& = &
c \Al \cdot \widehat{\vpi}\,,
\label{A-FW-2}
\eea
where
\bea
{\mbox{\boldmath $\alpha$}}
& = &
\left[
\ba{cc}
{\mbox{\boldmath $0$}} & {\mbox{\boldmath $\sigma$}} \nn \\
{\mbox{\boldmath $\sigma$}} & {\mbox{\boldmath $0$}}
\ea
\right]\,, \qquad
\beta
=
\left[
\ba{cc}
\one & {\mbox{\boldmath $0$}} \nn \\
{\mbox{\boldmath $0$}} & - \one
\ea
\right]\,, \qquad
\one
=
\left[
\ba{cc}
1 & 0 \nn \\
0 & 1
\ea
\right]\,, \nn \\
{\mbox{\boldmath $\sigma$}}
& = &
\left[
\sigma_x =
\left[
\ba{cc}
0 & 1 \\
1 & 0
\ea
\right]\,, \
\sigma_y =
\left[
\ba{lr}
0 & - \i \\
\i & 0
\ea
\right]\,, \
\sigma_z =
\left[
\ba{lr}
1 & 0 \\
0 & -1
\ea
\right]
\right].
\eea
with
$
\widehat{\vpi}
=
{\widehat{\mbox{\boldmath $p$}}} - q {\mbox{\boldmath $A$}}$,
$\widehat{\mbox{\boldmath $p$}} = - \i \hbar \Nab$, and
$\widehat{\pi}^2 =
\left(\widehat{\pi}_x^2 + \widehat{\pi}_y^2
+ \widehat{\pi}_z^2\right)$.

In the nonrelativistic situation the upper pair of components of the
Dirac Spinor $\Psi$ are large compared to the lower pair of components.
The operator $\widehat{\cal E}$ which does not couple the large and
small components of $\Psi$ is called `even' and $\widehat{\cal O}$ is
called an `odd' operator which couples the large to the small
components.  Note that
\beq
\beta \widehat{\cal O} = - \widehat{\cal O} \beta\,, \qquad
\beta \widehat{\cal E} = \widehat{\cal E} \beta\,.
\eeq
Now, the search is for a unitary transformation,
$\Psi'$ $=$ $\Psi$ $\longrightarrow$ $\widehat{U} \Psi$, such that the
equation for $\Psi '$ does not contain any odd operator.

In the free particle case (with $\phi = 0$
and $\widehat{\vpi} = \widehat{\p}$)
such a Foldy-Wouthuysen transformation is given by
\bea
\Psi \longrightarrow \Psi' & = & \widehat{U}_{F} \Psi \nn \\
\widehat{U}_F & = & e^{\i \widehat{S}} =
e^{\beta \Al \cdot \widehat{\p} \theta} \,,
\quad {\rm tan}\,2 | \widehat{\p} |
\theta = \frac{| \widehat{\p}|}{m_0 c}\,.
\eea
This transformation eliminates the odd part completely from the
free particle Dirac Hamiltonian reducing it to the diagonal form:
\bea
\i \hbar \frac{\partial}{\partial t} \Psi'
& = &
e^{\i \widehat{S}} \left({m_0 c^2} \beta
+ c \Al \cdot \widehat{\p} \right)
e^{- \i \widehat{S}} \Psi ' \nn \\
& = &
\left(\cos \,| \widehat{\p}| \theta +
\frac{\beta \Al \cdot \widehat{\p}}{| \widehat{\p} |} \sin \,|
\widehat{\p} | \theta \right)
\left({m_0 c^2} \beta + c \Al \cdot \widehat{\p} \right) \nn \\
& & \qquad \qquad
\times \left(\cos \,| \widehat{\p}| \theta -
\frac{\beta \Al \cdot \widehat{\p}}{| \widehat{\p} |} \sin \,|
\widehat{\p} | \theta \right) \Psi' \nn \\
& = &
\left(m_0 c^2 \cos \,2 | \widehat{\p}| \theta + c
| \widehat{\p} | \sin \,2
| \widehat{\p} | \theta \right) \beta \Psi' \nn \\
& = &
\left(\sqrt{m_0^2 c^4 + c^2 \widehat{p}^2} \right) \beta \,\Psi'\,.
\eea

In the general case, when the electron is in a time-dependent
electromagnetic field it is not possible to construct an
$\exp (\i \widehat{S})$ which removes the odd operators from the
transformed Hamiltonian completely.  Therefore, one has to be content
with a nonrelativistic expansion of the transformed Hamiltonian in a
power series in $1/{m_0 c^2}$ keeping through any desired order.  Note
that in the nonrelativistic case, when $|\p | \ll m_0 c$, the
transformation operator $\widehat{U}_F = \exp (\i \widehat{S})$ with
$\widehat{S} \approx - \i \beta \widehat{\cal O} /2 m_0 c^2$, where
$\widehat{\cal O} = c \Al \cdot \widehat{\p} $ is the odd part of the
free Hamiltonian.  So, in the general case we can start with the
transformation
\beq
\Psi^{(1)} = e^{\i \widehat{S}_1} \Psi, \qquad \widehat{S}_1 =
- \frac{\i \beta \widehat{\cal O}}{2 m_0 c^2}
= - \frac{\i \beta \Al \cdot \widehat{\vpi }}{2 m_0 c}\,.
\eeq
Then, the equation for $\Psi^{(1)}$ is
\bea
\i \hbar \frac{\partial}{\partial t} \Psi^{(1)}
& = &
\i \hbar \frac{\partial}{\partial t} \left(e^{\i \widehat{S}_1}
\Psi \right)
=
\i \hbar \frac{\partial}{\partial t} \left(e^{\i \widehat{S}_1}
\right) \Psi + e^{\i \widehat{S}_1} \left(\i \hbar
\frac{\partial}{\partial t} \Psi\right) \nn \\
& = &
\left[\i \hbar \frac{\partial}{\partial t}
\left(e^{\i \widehat{S}_1} \right) + e^{\i \widehat{S}_1}
\widehat{H}_D \right] \Psi \nn \\
& = &
\left[ \i \hbar \frac{\partial}{\partial t}
\left(e^{\i \widehat{S}_1} \right)
e^{- \i \widehat{S}_1}
+ e^{\i \widehat{S}_1} \widehat{H}_D e^{- \i \widehat{S}_1}
\right] \Psi^{(1)} \nn \\
& = &
\left[
e^{\i \widehat{S}_1} \widehat{H}_D e^{- \i \widehat{S}_1}
- \i \hbar e^{\i \widehat{S}_1}
\frac{\partial}{\partial t} \left(e^{- \i \widehat{S}_1} \right)
\right] \Psi^{(1)} \nn \\
& = &
\widehat{H}_D^{(1)} \Psi^{(1)}
\eea
where we have used the identity $\frac{\partial}{\partial t}
\left(e^{ \widehat{A}} \right) e^{- \widehat{A}}$ $+$ $e^{\widehat A}
\frac{\partial}{\partial t} \left(e^{ - \widehat{A}} \right)$
$ = $ $\frac{\partial}{\partial t} \widehat{I}$ $ = 0$.

Now, using the identities
\bea
e^{\widehat{A}} \widehat{B} e^{-\widehat{A}}
& = &
\widehat{B} + [\widehat{A} , \widehat{B} ]
+ \frac{1}{2!} [\widehat{A} , [ \widehat{A} , \widehat{B} ]]
+ \frac{1}{3!} [\widehat{A} ,
[ \widehat{A} , [ \widehat{A} , \widehat{B} ]]] + \ldots \nn \\
& &
e^{\widehat{A}(t)} \frac{\partial}{\partial t}
\left(e^{-\widehat{A}(t)} \right)
\nn \\
& & \ \
= \left( 1 + {\widehat{A}(t)} + \frac{1}{2!} {\widehat{A}(t)}^2
+ \frac{1}{3!} {\widehat{A}(t)}^3 \cdots \right) \nn \\
& & \ \ \quad \quad \times \frac{\partial}{\partial t}
\left( 1 - {\widehat{A}(t)} + \frac{1}{2!} {\widehat{A}(t)}^2
- \frac{1}{3!} {\widehat{A}(t)}^3 \cdots \right) \nn \\
& & \ \
= \left(1 + \At + \frac{1}{2!} \At^2
+ \frac{1}{3!} \At^3 \cdots \right) \nn \\
& & \ \ \quad \quad
\times \left(- \dAt + \frac{1}{2!} \left\{\dAt \At +
\At \dAt \right\} \right. \nn \\
& & \ \ \quad \quad
- \frac{1}{3!} \left\{\dAt \At^2 + \At \dAt \At \right.  \nn \\
& & \ \ \quad \quad \left. \left.
+ \At^2 \dAt \right\} \ldots \right) \nn \\
& & \ \ \approx
- \dAt - \frac{1}{2!} \left[\At , \dAt \right] \nn \\
& & \ \ \quad \quad
- \frac{1}{3!} \left[\At , \left[\At , \dAt \right] \right] \nn \\
& & \ \ \quad \quad
- \frac{1}{4!} \left[\At , \left[ \At , \left[ \At , \dAt
\right] \right] \right]\,,
\eea
with $\widehat{A} = {\i \widehat{S}_1}$, we find
%
\bea
\widehat{H}_D^{(1)} & \approx & \widehat{H}_D - \hbar \dsone
+ \i \left[\sone , \widehat{H}_D - \frac{\hbar}{2} \dsone \right] \nn \\
& & \qquad
- \frac{1}{2!} \left[ \sone , \left[\sone ,
\widehat{H}_D - \frac{\hbar}{3} \dsone \right] \right] \nn \\
& & \qquad
- \frac{\i}{3!} \left[ \sone , \left[\sone , \left[\sone ,
\widehat{H}_D - \frac{\hbar}{4} \dsone \right] \right] \right]\,.
\label{A-FW-9}
\eea
Substituting in~(\ref{A-FW-9}),
$\widehat{H}_D = {m_0 c^2} \beta + \widehat{\cal E}
+ \widehat{\cal O}$, simplifying the right hand side using the
relations $\beta \widehat{\cal O} = - \widehat{\cal O} \beta$ and
$\beta \widehat{\cal E} = \widehat{\cal E} \beta$ and collecting
everything together, we have
\bea
\widehat{H}_D^{(1)}
& \approx &
{m_0 c^2} \beta + \widehat{\cal E}_1 + \widehat{\cal O}_1 \nn \\
\widehat{\cal E}_1
& \approx &
\E + \frac{1}{2 m_0 c^2} \beta \O^2 - \frac{1}{8 m_0^2 c^4}
\left[\O ,
\left( \left[\O , \E \right] +  \i \hbar \dO \right) \right] \nn \\
& & \quad
- \frac{1}{8 m_0^3 c^6} \beta \O^4 \nn \\
\O_1
& \approx & \frac{\beta}{2 m_0 c^2}
\left(\left[\O , \E \right] + \i \hbar \dO \right)
- \frac{1}{3 m_0^2 c^4} \O^3\,,
\eea
with $\E_1$ and $\O_1$ obeying the relations
$\beta \widehat{\cal O}_1 = - \widehat{\cal O}_1 \beta$
and $\beta \widehat{\cal E}_1 = \widehat{\cal E}_1 \beta$
exactly like $\E$ and $\O$.  It is seen that while the term $\O$ in
$\widehat{H}_D$ is of order zero with respect to the expansion parameter
$1/{m_0 c^2}$
({\em i.e.}, $\O$ $=$ $O \left( \left( 1/{m_0 c^2} \right)^0 \right)$
the odd part of
$\widehat{H}_D^{(1)} $, namely $\O_1$, contains only terms of
order $1/{m_0 c^2}$ and higher powers of $1/{m_0 c^2}$
({\em i.e.}, $\O_1 = O \left( \left(1/{m_0 c^2}\right) \right)$).

To reduce the strength of the odd terms further in the transformed
Hamiltonian a second Foldy-Wouthuysen transformation is applied with
the same prescription:
\bea
\Psi^{(2)}
& = &
e^{\i \widehat{S}_2} \Psi^{(1)} \,, \nn \\
\qquad \widehat{S}_2
& = &
- \frac{ \i \beta \widehat{\cal O}_1}{2 m_0 c^2} \nn \\
& = &
- \frac{\i \beta}{2 m_0 c^2} \left[
\frac{\beta}{2 m_0 c^2}
\left( \left[\O , \E \right] + \i \hbar \dO \right)
- \frac{1}{3 m_0^2 c^4} \O^3 \right]\,. 
\eea
After this transformation,
\bea
\i \hbar \frac{\partial}{\partial t} \Psi^{(2)}
& = &
\widehat{H}_D^{(2)} \Psi^{(2)}\,, \quad
\widehat{H}_D^{(2)}
=
{m_0 c^2} \beta + \widehat{\cal E}_2 + \widehat{\cal O}_2 \nn \\
\widehat{\cal E}_2 wide
& \approx &
\E_1\,, \quad
\O_2 \approx \frac{\beta}{2 m_0 c^2}
\left(\left[\O_1 , \E_1 \right] + \i \hbar
\frac{\partial \O_1}{\partial t} \right)\,, 
\eea
where, now, $\O_2 = O \left(\left(1/{m_0 c^2}\right)^2 \right)$.
After the third transformation
\beq
\Psi^{(3)} = e^{\i \widehat{S}_3}\,\Psi^{(2)}, \qquad \widehat{S}_3 =
- \frac{ \i \beta \widehat{\cal O}_2}{2 m_0 c^2} \nn
\eeq
we have
\bea
\i \hbar \frac{\partial}{\partial t} \Psi^{(3)}
& = &
\widehat{H}_D^{(3)} \Psi^{(3)}\,, \quad
\widehat{H}_D^{(3)}
=
{m_0 c^2} \beta + \widehat{\cal E}_3 + \widehat{\cal O}_3 \nn \\
\widehat{\cal E}_3 & \approx & \E_2 \approx \E_1\,, \quad
\O_3 \approx \frac{\beta}{2 m_0 c^2} \left(\left[\O_2 , \E_2 \right]
+ \i \hbar \frac{\partial \O_2}{\partial t} \right)\,, 
\eea
where $\O_3 = O \left( \left( 1/{m_0 c^2} \right)^3 \right)$. So,
neglecting $\O_3$,
\bea
\widehat{H}_D^{(3)}
& \approx &
{m_0 c^2} \beta + \widehat{\cal E} +
\frac{1}{2 m_0 c^2} \beta \widehat{\cal O}^2 \nn \\
& & \quad
- \frac{1}{8 m_0^2 c^4} \left[\O , \left( \left[\O , \E \right]
+ \i \hbar \frac{\partial \O }{\partial t} \right) \right] \nn \\
& & \quad
-
\frac{1}{8 m_0^3 c^6} \beta
\left\{
\O^4
+
\left(\left[\O , \E \right] + \i \hbar
\frac{\partial \O }{\partial t} \right)^2
\right\}
\label{A-FW-FOUR}
\eea
It may be noted that starting with the second transformation
successive $(\E , \O)$ pairs can be obtained recursively using the
rule
\bea
\E_j & = & \E_1 \left(\E \rightarrow \E_{j-1} ,
\O \rightarrow \O_{j-1} \right) \nn \\
\O_j
& = &
\O_1 \left(\E \rightarrow \E_{j-1} ,
\O \rightarrow \O_{j-1} \right)\,, \quad j > 1\,,
\eea
and retaining only the relevant terms of desired order at each step.

With $\widehat{\cal E} = q \phi$ and
$\widehat{\cal O} = c \Al \cdot \widehat{\vpi}$, the final reduced
Hamiltonian~(\ref{A-FW-FOUR}) is, to the order calculated,
\bea
\widehat{H}_D^{(3)}
& = &
\beta \left({m_0 c^2} + \frac{\widehat{\pi}^2}{2 m_0}
- \frac{\widehat{p}^4}{8 m_0^3 c^6} \right) + q \phi
- \frac{ q \hbar}{2 m_0 c} \beta \Vsig \cdot \B \nn \\
& & \quad
- \frac{\i q {\hbar}^2}{8 m_0^2 c^2} \Vsig \cdot
{\rm curl}\,{\mbox{\boldmath $E$}}
- \frac{q{\hbar}}{4 m_0^2 c^2} \Vsig \cdot
{\mbox{\boldmath $E$}} \times \widehat{\p} \nn \\
& & \quad
- \frac{q{\hbar}^2}{8 m_0^2 c^2}
{\rm div}{\mbox{\boldmath $E$}}\,,
\eea
with the individual terms having direct physical interpretations. The
terms in the first parenthesis result from the expansion of
$\sqrt{m_0^2 c^4 + c^2 \widehat{\pi}^2}$
showing the effect of the relativistic mass increase. The second and
third terms are the electrostatic and magnetic dipole energies. The
next two terms, taken together (for hermiticity), contain the
spin-orbit interaction. The last term, the so-called Darwin term,
is attributed to the {\em zitterbewegung} (trembling motion) of the
Dirac particle: because of the rapid coordinate fluctuations over
distances of the order of the Compton wavelength ($2 \pi \hbar /m_0 c$)
the particle sees a somewhat smeared out electric potential.

It is clear that the Foldy-Wouthuysen transformation technique expands
the Dirac Hamiltonian as a power series in the parameter
$1/{m_0 c^2}$ enabling the use of a systematic approximation
procedure for studying the deviations from the nonrelativistic
situation.  We note the analogy between the nonrelativistic
particle dynamics and paraxial optics:

\begin{center}
{\bf The Analogy}
\end{center}
\begin{tabular}{ll}
{\bf Standard Dirac Equation} ~~~~~~~ & {\bf Beam Optical Form} \\
$m_0 c^2 \beta + \E_D + \O_D$ & $- n_0 \beta + \E + \O$ \\
$m_0 c^2$ & $- n_0$ \\
Positive Energy & Forward Propagation \\
Nonrelativistic, $\left|\mbox{\boldmath $\pi$}\right| \ll m_0 c$ &
Paraxial Beam, $\left|\hatp_\perp \right| \ll n_0$ \\
Non relativistic Motion  & Paraxial Behavior \\
~~ + Relativistic Corrections & ~~ + Aberration Corrections \\
\end{tabular}

\bigskip

Noting the above analogy, the idea of Foldy-Wouthuysen form of the
Dirac theory has been adopted to study the paraxial optics and
deviations from it by first casting the Maxwell equations in a spinor
form resembling exactly the Dirac equation~(\ref{A-FW-1}, \ref{A-FW-2})
in all respects: {\em i.e}., a multicomponent $\Psi$ having the upper
half of its components large compared to the lower components and the
Hamiltonian having an even part $(\E)$, an odd part $(\O)$, a suitable
expansion parameter, ($|\hatp_\perp|/{n_0} \ll 1$) characterizing the
dominant forward
propagation and a leading term with a $\beta$ coefficient commuting
with $\E$ and anticommuting with $\O$.  The additional feature of
our formalism is to return finally to the original representation
after making an extra approximation, dropping $\beta$ from the final
reduced optical Hamiltonian, taking into account the fact that we are
primarily interested only in the forward-propagating beam.

\newpage

\section*{}
\addcontentsline{toc}{section}
{Appendix D. \\
The Magnus Formula}

\renewcommand{\theequation}{D.{\arabic{equation}}}
\setcounter{equation}{0}

\begin{center}

{\Large\bf
Appendix-D \\
The Magnus Formula
} \\

\end{center}

The Magnus formula is the continuous analogue of the famous
Baker-Campbell-Hausdorff (BCH) formula
\beq
\e ^{{\hat A}} \e ^{{\hat B}} =
\e ^{ {\hat A} +  {\hat B} +
\half [ {\hat A}, {\hat B}] + \frac{1}{12} \left\{
[ [ {\hat A}, {\hat A}], {\hat B} ] +
[ [ {\hat A}, {\hat B}], {\hat B} ] \right\} + \ldots }\,.
\label{BCH}
\eeq
Let it be required to solve the differential equation
\beq
\frac{\partial}{\partial t} u(t) = {\hat A} (t) u(t)
\label{Magnus-2}
\eeq
to get $u(T)$ at $T > t_0$, given the value of $u (t_0)$; the
operator ${\hat A}$ can represent any linear operation.  For an
infinitesimal $ \Delta t$, we can write
\beq
u (t_0 + \Delta t) = e^{\Delta t {\hat A}(t_0)} u (t_0).
\eeq
Iterating this solution we have
\bea
u(t_0 + 2 \Delta t) & = & \e ^ {\Delta t {\hat A}(t_0 + \Delta t)}
\e ^ {\Delta t {\hat A}(t_0 )} u(t_0)
\nn \\
u(t_0 + 3 \Delta t) & = & \e ^ {\Delta t {\hat A}(t_0 + 2 \Delta t)}
\e ^ {\Delta t {\hat A}(t_0 + \Delta t)}
\e ^ {\Delta t {\hat A}(t_0 )} u(t_0)
\nn \\
&   & \ldots \quad {\rm and\ so\ on}.
\eea
If $ T = t_0 + N \Delta t$ we would have
\beq
u(T) = \left\{ \prod_{n =0}^{N-1}
e ^ {\Delta t {\hat A}(t_0 + n \Delta t)} \right\} u(t_0)\,.
\label{pi}
\eeq
Thus, $u(T)$ is given by computing the product in~(\ref{pi})
using
successively the BCH-formula~(\ref{BCH}) and considering the limit $\Delta t
\longrightarrow 0, N \longrightarrow \infty $ such that
$N \Delta t = T -t_0 $. The resulting expression is the Magnus
formula~(Magnus,~\cite{Magnus})~:
\bea
u(T) & = &  \widehat{\cal T}(T, t_0) u(t_0)
\nn \\
{\cal T}(T, t_0) & = & \exp \left\{
\int_{t_0}^{T} d t_1\,{\widehat A}(t_1) \right.
\nn \\
&   & \ + \half \int_{t_0}^{T} d t_2
\int_{t_0}^{t_2} d t_1\,\left[ {\hat A}(t_2), {\hat A}(t_1) \right]
\nn \\
&  & \ + \frac{1}{6}
\int_{t_0}^{T} d t_3 \int_{t_0}^{t_3} d t_2 \int_{t_0}^{t_2} d t_1\,\left(
\left[ \left[ {\hat A}(t_3), {\hat A}(t_2)\right], {\hat A}(t_1) \right] \right.
\nn \\
&  & \left. \phantom{\int_{t_0}^T} {\left. \quad \quad  + \left[ \left[ {\hat
A}(t_1), {\hat A}(t_2) \right], {\hat A}(t_3) \right] \right)} +\,\ldots
\right\}\,.
\label{Magnus-6}
\eea

To see how the equation~(\ref{Magnus-6}) is obtained let us substitute
the assumed form of the solution,
$u (t) = \ct \left(t , t_0 \right) u \left(t_0 \right)$,
in~(\ref{Magnus-2}). Then, it is seen that
$\widehat{{\cal T}} (t, t_0)$
obeys the equation
\beq
\frac{\partial }{\partial t} \widehat{{\cal T}} (t, t_0) =
{\widehat A} (t) {\cal T} (t, t_0),
\qquad {\widehat{\cal T}} (t_0, t_0) = {\hat I}\,.
\label{Magnus-7}
\eeq
Introducing an iteration parameter $\lambda$
in~(\ref{Magnus-7}), let
\bea
\frac{\partial}{\partial t} \widehat{{\cal T}}(t, t_0; \lambda)
& = &
\lambda {\widehat A}(t) \widehat{{\cal T}}(t, t_0; \lambda)\,,
\label{Magnus-8}
\\
\widehat{{\cal T}}(t_0, t_0; \lambda) & = &  {\widehat I}\,, \quad
\widehat{{\cal T}}(t, t_0; 1) = \widehat{{\cal T}}(t, t_0)\,.
\label{Magnus-9}
\eea
Assume a solution of~(A8) to be of the form
\beq
\widehat{{\cal T}}(t, t_0; \lambda)
= \e ^{{\Omega} (t, t_0; \lambda)}
\eeq
with
\beq
{\Omega}(t, t_0; \lambda) = \sum_{n =1}^{\infty} {\lambda} ^n
\Delta_n (t, t_0), \quad \Delta_n (t_0, t_0) = 0 \quad
{\rm for \ all \ } n\,.
\label{Magnus-11}
\eeq
Now, using the identity (se Wilcox,~\cite{Wilcox})
\beq
\frac{\partial}{\partial t} \e ^{ {\Omega}(t, t_0; \lambda)} =
\left\{
\int_{0}^{1} ds e ^{s \Omega (t, t_0; \lambda)}
\frac{\partial}{\partial t} {\Omega}(t, t_0; \lambda)
\e ^{ - s \Omega (t, t_0; \lambda)}  \right\} \e ^{{\Omega}(t, \lambda)}\,,
\eeq
one has
\beq
\int_{0}^{1} ds e ^{s \Omega (t, t_0; \lambda)}
\frac{\partial}{\partial t} {\Omega}(t, t_0; \lambda)
e ^{ - s \Omega (t, t_0; \lambda)} =
\lambda \widehat{A} (t)\,.
\label{Magnus-13}
\eeq
Substituting in~(\ref{Magnus-13}) the series expression for
$\Omega (t, t_0; \lambda)$~(\ref{Magnus-11}), expanding the left hand
side using the first identity in~(\ref{Magnus-8}), integrating and
equating the coefficients of $ \lambda ^j$ on both sides, we get,
recursively, the equations for
$ \Delta_1 (t, t_0)$, $\Delta_2 (t, t_0), \ldots \,,$ etc. For $j = 1$
\beq
\frac{\partial}{\partial t} \Delta_1 (t, t_0) = {\hat A}(t),
\qquad \Delta_1 (t_0, t_0) = 0
\eeq
and hence
\beq
\Delta_1 (t, t_0) =  \int_{t_0}^{t} d t_1 \widehat{A} (t_1)\,.
\eeq
For $j=2$
\beq
\frac{\partial}{\partial t} \Delta_2 (t, t_0) +
\half \left[ \Delta_1 (t, t_0)\,,\,\frac{\partial}{\partial t}
\Delta_1 (t, t_0) \right] =  0\,,
\qquad \Delta_2 (t_0, t_0) = 0
\eeq
and hence
\beq
\Delta_2 (t, t_0) =  \half
\int_{t_0}^{t} d t_2
\int_{t_0}^{t_2} d t_1 \left[ \widehat{A} (t_2)\,,\,\widehat{A}
(t_1) \right].
\eeq
Similarly,
\bea
\Delta_3 (t, t_0) & = & \frac{1}{6}
\int_{t_0}^{t} d t_1 \int_{t_0}^{t_1} d t_2 \int_{t_0}^{t_2} d t_3\,\left\{
\left[ \left[ {\hat A}(t_1)\,,\,{\hat A}(t_2) \right]\,,\,{\hat A}(t_3) \right]
\right.
\nn \\
&   & \quad \quad \left. + \left[ \left[ {\hat A}(t_3)\,,\,{\hat A}(t_2)
\right]\,,\,{\hat A}(t_1) \right] \right\}\,.
\eea
Then, the Magnus formula in~(\ref{Magnus-6}) follows
from~(\ref{Magnus-9})-(\ref{Magnus-11}).  Equation~(\ref{T-Magnus}) we
have, in the context of $z$-evolution follows from the above discussion
with the identification
$t \longrightarrow z$, $t_0 \longrightarrow z^{(1)}$,
$T \longrightarrow z^{(2)}$
and ${\hat A}(t) \longrightarrow - \ih \ho (z)$.

For more details on the exponential solutions of linear differential
equations, related operator techniques and applications to
physical problems the reader is referred to Wilcox~\cite{Wilcox},
Bellman and Vasudevan~\cite{BV}, Dattoli {\em et al.}~\cite{DRT},
and references therein.

\newpage

\section*{}
\addcontentsline{toc}{section}
{Appendix E. \\
Analogies between light optics and charged-particle optics:
Recent Developments
}

%
\renewcommand{\theequation}{E.{\arabic{equation}}}
\renewcommand{\thesection}{E.{\arabic{section}}}
\renewcommand{\thesubsection}{E.{\arabic{subsection}}}
\setcounter{subsection}{0}
\setcounter{equation}{0}

\begin{center}

{\Large\bf
Appendix-E \\
Analogies between light optics and charged-particle optics:
Recent Developments
} \\

\end{center}

Historically, variational principles have played a fundamental role
in the evolution of mathematical models in classical physics, and many
equations can be derived by using them.  Here the relevant examples are
Fermat's principle in optics and Maupertuis' principle in mechanics.
The beginning of the analogy between geometrical optics and mechanics
is usually attributed to Descartes (1637), but actually it can traced
back to Ibn Al-Haitham Alhazen (0965-1037)~\cite{Ambrosini}.  The
analogy between the trajectory of material particles in potential
fields and the path of light rays in media with continuously variable
refractive index was formalized by Hamilton in 1833.  This Hamiltonian
analogy lead to the development of electron optics in 1920s, when Busch
derived the focusing action and a lens-like action of the axially
symmetric magnetic field using the methodology of geometrical optics.
Around the same time Louis de Broglie associated his now famous
wavelength to moving particles.  Schr\"{o}dinger extended the analogy
by passing from geometrical optics to wave optics through his wave
equation incorporating the de Broglie wavelength.  This analogy played
a fundamental role in the early development of quantum mechanics.  The
analogy, on the other hand, lead to the development of practical
electron optics and one of the early inventions was the electron
microscope by Ernst Ruska.  A detailed account of Hamilton's analogy
is available in~\cite{Hawkes}-\cite{Forbes}.

Until very recently, it was possible to see this analogy only between
the geometrical-optic and classical prescriptions of electron optics.
The reasons being that, the quantum theories of charged-particle beam
optics have been under development only for about a
decade~\cite{JSSM}-\cite{JK} with the very expected feature of
wavelength-dependent effects, which have no analogue in the traditional
descriptions of light beam optics.  With the  current development of
the non-traditional prescriptions of Helmholtz
optics~\cite{KJS-1,Khan-H-1} and the matrix formulation of Maxwell
optics, accompanied with
wavelength-dependent effects, it is seen that the analogy between the
two systems persists.  The non-traditional prescription of Helmholtz
optics is in close analogy with the quantum theory of charged-particle
beam optics based on the Klein-Gordon equation.  The matrix formulation
of Maxwell optics is in close analogy with the quantum theory of
charged-particle beam optics based on the Dirac equation.  This analogy
is summarized in the table of Hamiltonians.  In this short note it is
difficult to present the derivation of the various Hamiltonians which
are available in the references.  We shall briefly consider an outline
of the quantum prescriptions and the non-traditional prescriptions
respectively.  A complete coverage to the new field of  {\em Quantum
Aspects of Beam Physics} ({\bf QABP}), can be found in the proceedings
of the series of meetings under the same name~\cite{QABP}.

\subsection{Quantum Formalism of Charged-Particle Beam Optics}
The classical treatment of charged-particle beam optics has been
extremely successful in the designing and working of numerous optical
devices, from electron microscopes to very large particle accelerators.
It is natural, however to look for a prescription based on the quantum
theory, since any physical system is quantum mechanical at the
fundamental level!  Such a prescription is sure to explain the grand
success of the classical theories.  It is sure to help get a deeper
understanding and lead to better designing of charged-particle beam
devices.

The starting point of the quantum prescription of charged particle
beam optics is to build a theory based on the basic equations of
quantum mechanics (Schr\"{o}dinger, Klein-Gordon, Dirac) appropriate
to the situation under study.  In order to analyze the evolution of the
beam parameters of the various individual beam optical elements
(quadrupoles, bending magnets,~$\cdots$) along the optic axis of the
system, the first step is to start with the basic time-dependent
equations of quantum mechanics and then obtain an equation of the form
\begin{equation}
\i \hbar \frac{\partial }{\partial s} \psi \left(x , y ;\, s \right)
=
\widehat{\cal H} \left(x , y ;\, s \right)
\psi \left(x , y ;\, s \right)\,,
\label{BOE}
\end{equation}
where $(x , y ;\, s)$ constitute a curvilinear coordinate
system, adapted to the geometry of the system.  Eq.~(\ref{BOE}) is
the basic equation in the quantum formalism, called as the
{\em beam-optical equation}; ${\cal H}$ and $\psi$ as the
{\em beam-optical Hamiltonian} and the {\em beam wavefunction}
respectively.  The second step requires obtaining a relationship
between any relevant observable $\{\langle O \rangle (s) \}$ at the
transverse-plane at $s$ and the observable
$\{\langle O \rangle (s_{\rm in}) \}$
at the transverse plane at $s _{\rm in}$, where $s _{\rm in}$ is some
input reference point.  This is achieved by the integration of the
beam-optical equation in~(\ref{BOE})
\begin{eqnarray}
\psi \left(x , y ; s \right) & = &
\widehat{U} \left(s , s_{\rm in} \right)
\psi \left(x , y ; s_{\rm in} \right)\,,
\label{BOI}
\end{eqnarray}
which gives the required transfer maps
\begin{eqnarray}
\left\langle O \right\rangle \left(s_{\rm in} \right)
\longrightarrow
\left\langle O \right\rangle \left(s \right)
& = &
\left\langle \psi \left(x , y ; s \right)
\left| O \right|
\psi \left(x , y ; s \right) \right\rangle\,, \nn \\
& = &
\left\langle \psi \left(x , y ; s_{\rm in} \right)
\left| \widehat{U} ^{\dagger} O  \widehat{U} \right|
\psi \left(x , y ; s_{\rm in} \right) \right\rangle\,.
\label{BOM}
\end{eqnarray}

The two-step algorithm stated above gives an over-simplified picture of
the quantum formalism.  There are several crucial points to be noted.
The first step in the algorithm of obtaining the beam-optical equation
is not to be treated as a mere transformation which eliminates $t$ in
preference to a variable $s$ along the optic axis.  A clever set of
transforms are required which not only eliminate the variable $t$ in
preference to $s$ but also give us the $s$-dependent equation which has
a close physical and mathematical correspondence with the original
$t$-dependent equation of standard time-dependent quantum mechanics.
The imposition of this stringent requirement on the construction of the
beam-optical equation ensures the execution of the second-step of the
algorithm.  The beam-optical equation is such that all the required
rich machinery of quantum mechanics becomes applicable to the
computation of the transfer maps that characterize the optical system.
This describes the essential scheme of obtaining the quantum formalism.
The rest is mostly mathematical detail which is inbuilt in the powerful
algebraic machinery of the algorithm, accompanied with some reasonable
assumptions and approximations dictated by the physical considerations.
The nature of these approximations can be best summarized in the optical
terminology as a systematic procedure of expanding the beam optical
Hamiltonian in a power series of $|{\widehat{\vpi}_\perp}/{p_0}|$,
where $p_0$ is the design (or average) momentum of beam particles
moving predominantly along the direction of the optic axis and
$\widehat{\vpi}_\perp$ is the small transverse kinetic momentum.  The
leading order approximation along with
$|{\widehat{\vpi}_\perp}/{p_0}| \ll 1$, constitutes the paraxial or
ideal behaviour and higher order terms in the expansion
give rise to the nonlinear or aberrating behaviour.
It is seen that the paraxial and aberrating behaviour get modified by
the quantum contributions which are in powers of the de Broglie
wavelength ($\LAMBDA_0 = {\hbar}/{p_0}$).  The classical limit
of the quantum formalism reproduces the well known Lie algebraic
formalism of charged-particle beam optics~\cite{Lie}.

\subsection{Light Optics: Various Prescriptions}
The traditional scalar wave theory of optics (including aberrations to
all orders) is based on the beam-optical Hamiltonian derived by using
Fermat's principle.  This approach is purely geometrical and works
adequately in the scalar regime.  The other approach is based on the
{\em square-root} of the Helmholtz operator, which is derived from the
Maxwell equations~\cite{Lie}.  This approach works to all orders and
the resulting expansion is no different from the one obtained using
the geometrical approach of Fermat's principle.  As for the
polarization: a systematic procedure for the passage from scalar to
vector wave optics to handle paraxial beam propagation problems,
completely taking into account the way in which the Maxwell equations
couple the spatial variation and polarization of light waves, has been
formulated by analyzing the basic Poincar\'{e} invariance of the system,
and this procedure has been successfully used to clarify several issues
in Maxwell optics~\cite{MSS-1}-\cite{SSM-2}.

In the above approaches, the beam-optics and the polarization are
studied separately, using very different machineries.  The derivation
of the Helmholtz equation from the Maxwell equations is an
approximation as one neglects the spatial and temporal derivatives of
the permittivity and permeability of the medium.  Any prescription
based on the Helmholtz equation is bound to be an approximation,
irrespective of how good it may be in certain situations.  It is very
natural to look for a prescription based fully on the Maxwell
equations, which is sure to provide a deeper
understanding of beam-optics and light polarization in a unified manner.

The two-step algorithm used in the construction of the quantum theories
of charged-particle beam optics is very much applicable in light optics!
But there are some very significant conceptual differences to be borne
in mind.  When going beyond Fermat's principle the whole of optics
is completely governed by the Maxwell equations, and there are no other
equations, unlike in quantum mechanics, where there are separate
equations for, spin-$1/2$, spin-$1$, $\cdots$.

Maxwell's equations are linear (in time and space derivatives) but
coupled in the fields.  The decoupling leads to the Helmholtz equation
which is quadratic in derivatives.  In the specific context of beam
optics, purely from a calculational point of view, the starting
equations are the Helmholtz equation governing scalar optics and for a
more accurate prescription one uses the full set of Maxwell equations,
leading to vector optics.  In the context of the two-step algorithm,
the Helmholtz equation and the Maxwell equations in a matrix
representation can be treated as the `basic' equations, analogue of
the basic equations of quantum mechanics.  This works perfectly fine
from a calculational point of view in the scheme of the algorithm we
have.

Exploiting the similarity between the Helmholtz wave equation and the
Klein-Gordon equation, the former is linearized using
the Feshbach-Villars procedure used for the linearization of the
Klein-Gordon equation.  Then the Foldy-Wouthuysen iterative
diagonalization technique is applied to obtain a Hamiltonian description
for a system with varying refractive index.  This technique is an
alternative to the conventional method of series expansion of the
radical.  Besides reproducing all the traditional quasiparaxial terms,
this method leads to additional terms, which are dependent on the
wavelength, in the optical Hamiltonian.
This is the non-traditional prescription of scalar optics.

The Maxwell equations can be cast into an exact matrix form taking into
account the spatial and temporal variations of the permittivity and
permeability.  The derived  representation using $8 \times 8$ matrices
has a close algebraic analogy with the Dirac equation, enabling the use
of the rich machinery of the Dirac electron theory.  The beam optical
Hamiltonian derived from this representation reproduces the
Hamiltonians obtained in the traditional prescription along with
wavelength-dependent matrix terms, which we have named as the
{\em polarization terms}.  These polarization terms are very similar
to the spin terms in the Dirac electron theory and the spin-precession
terms in the beam-optical version of the Thomas-BMT
equation~\cite{CJKP-1}.  The
matrix formulation provides a unified treatment of beam optics and light
polarization.  Some well known results of light polarization are
obtained as the paraxial limit of the matrix
formulation~\cite{MSS-1}-\cite{SSM-2}.
The traditional beam optics is completely obtained from our approach
in the limit of small wavelength, $\LAMBDA \longrightarrow 0$, which
we call as the traditional limit of our formalisms.  This is analogous
to the classical limit obtained by taking $\hbar \longrightarrow 0$,
in the quantum prescriptions.

From the Hamiltonians in the Table we make the following observations:
The classical/traditional Hamiltonians of particle/light optics are
modified by wavelength-dependent contributions in the
quantum/non-traditional prescriptions respectively.  The algebraic
forms of these modifications in each row is very similar.  This should
not come as a big surprise.  The starting equations have one-to-one
algebraic correspondence: Helmholtz $\leftrightarrow$ Klein-Gordon;
Matrix form of Maxwell $\leftrightarrow$ Dirac equation.  Lastly, the
de Broglie wavelength, ${\LAMBDA}_0$, and $\LAMBDA$ have an
analogous status, and the classical/traditional limit is obtained by
taking ${\LAMBDA}_0 \longrightarrow 0$ and
$\LAMBDA \longrightarrow 0$ respectively.  The parallel of the
analogies between the two systems is sure to provide us with more
insights.

\section*{}
\addcontentsline{toc}{section}
{Appendix F. \\
An Invitation to the Experimentalists}

\renewcommand{\theequation}{F.{\arabic{equation}}}
\setcounter{equation}{0}

\begin{center}

{\Large\bf
Appendix-F. \\
An Invitation to the Experimentalists
} \\

\end{center}

It would be worthwhile to experimentally look for the predicted
{\it image rotation} and the wavelength-dependent modifications of
the aberration coefficients.

\newpage

\section*{}
\addcontentsline{toc}{section}
{Table A. \\
Hamiltonians in Different Prescriptions}

\begin{center}

{\Large\bf
Table A. \\
Hamiltonians in Different Prescriptions
} \\
\end{center}

{\small
\noindent
The following are the Hamiltonians, in the different prescriptions
of light beam optics and charged-particle beam optics for magnetic
systems.  $\widehat{H}_{0\,, p}$ are the paraxial Hamiltonians, with
lowest order wavelength-dependent contributions.

\noindent
\begin{tabular*}{6.0in}[t]{@{\extracolsep{\fill}}|ll|}
\hline
& \\
\parbox[t]{2.5in}{
\parbox[t]{2.5in}{
{\large\bf Light Beam Optics}
}}
&
\parbox[t]{3.0in}{
{\large\bf Charged-Particle Beam Optics}
} \\
& \\
\hline
& \\
\parbox[t]{2.5in}{
{\bf Fermat's Principle} \\

$
{\cal H}
=
- \left\{n^2 (\r) - \p_\perp^2 \right\}^{1/2}
$
}

&

\parbox[t]{2.5in}{
{\bf Maupertuis' Principle} \\

$
{\cal H}
=
- \left\{p_0^2 - {\vpi}_\perp^2 \right\}^{1/2} - q A_z
$
} \\
& \\
\hline
& \\
\parbox[t]{2.5in}{

{\bf Non-Traditional Helmholtz} \\

$
\widehat{H}_{0\,, p} = \\
- n (\r)
+ \frac{1}{2 n_0} \hatp_{\perp}^2 \\
- \frac{\i \LAMBDA}{16 n_0^3}
\left[\hatp_\perp^2 , \ddz n (\r) \right]
$

}

&

\parbox[t]{2.5in}{
{\bf Klein-Gordon Formalism} \\

$
\widehat{H}_{0\,, p} = \\
- p_0 - q A_z + \frac{1}{2 p_0} \widehat{\vpi}_\perp^2 \\
+ \frac{\i \hbar}{16 p_0^4}
\left[\widehat{\vpi}_\perp^2 \,, \ddz \widehat{\vpi}_\perp^2 \right]
$
} \\
& \\
\hline
& \\

\parbox[t]{2.5in}{

{\bf Maxwell, Matrix} \\

$
\widehat{H}_{0\,, p} = \\
- n (\r) + \frac{1}{2 n_0} \hatp_\perp^2 \\
- \i \LAMBDA \beta
{\mbox{\boldmath $\Sigma$}} \cdot {\mbox{\boldmath $u$}} \\
+
\frac{1}{2 n_0} \LAMBDA^2 w^2 \beta
$
}

&

\parbox[t]{2.5in}{
{\bf Dirac Formalism} \\

$
\widehat{H}_{0\,, p} = \\
- p_0 - q A_z  + \frac{1}{2 p_0} \widehat{\vpi}_\perp^2 \\
- \frac{\hbar}{2 p_0}
\left\{\mu \gamma
{\mbox{\boldmath $\Sigma$}}_\perp \cdot \B_\perp
+
\left(q + \mu \right) \Sigma_z B_z \right\} \\
+ \i \frac{\hbar}{m_0 c} \epsilon B_z
$
} \\
& \\
\hline
\end{tabular*}

\bigskip

\noindent
{\bf Notation}

\noindent
\begin{tabular*}{6.0in}[t]{@{\extracolsep{\fill}}ll}
\parbox[t]{2.5in}{
\parbox[t]{2.5in}{
$
{\rm Refractive ~ Index}, ~
n (\r) = c \sqrt{\epsilon (\r) \mu (\r)} \\
{\rm Resistance}, ~
h (\r) = \sqrt{{\mu (\r)}/{\epsilon (\r)}} \\
{\mbox{\boldmath $u$}} (\r)
=
- \frac{1}{2 n (\r)} \Nab n (\r) \\
{\mbox{\boldmath $w$}} (\r)
=
\frac{1}{2 h (\r)} \Nab h (\r) \\
$
${\mbox{\boldmath $\Sigma$}}$ and $\beta$ are the Dirac matrices.
}}

&

\parbox[t]{2.5in}{
$
\widehat{\vpi}_\perp
= {\widehat{\mbox{\boldmath $p$}}}_\perp
- q {\mbox{\boldmath $A$}}_\perp \\
\mu_a ~ {\rm anomalous ~ magnetic ~ moment}. \\
\epsilon_a ~ {\rm anomalous ~ electric ~ moment}. \\
\mu = {2 m_0 \mu_a}/{\hbar}\,, ~~~~
\epsilon = {2 m_0 \epsilon_a}/{\hbar} \\
\gamma = {E}/{m_0 c^2}
$
}
\end{tabular*}

}

\newpage

\addcontentsline{toc}{section}
{Bibliography}

\end{document}